\documentclass[preprintnumbers,amsmath,amssymb]{revtex4}
\usepackage{amsfonts}
\usepackage{graphics}
\usepackage{epsfig}
\usepackage{url}
\usepackage{multirow}
\newcommand {\be}{\begin{equation}}
\newcommand {\ee}{\end{equation}}
\newcommand {\ba}{\begin{eqnarray}}
\newcommand {\ea}{\end{eqnarray}}

\begin{document}
\title{SLIM at LHC:
LHC search power for a model linking \\
dark matter and neutrino mass }
\author{Y. Farzan }

\affiliation{School of physics, Institute for Research in
Fundamental Sciences (IPM), P.O. Box 19395-5531, Tehran, IRAN}
\email{yasaman@theory.ipm.ac.ir}
\author{M. Hashemi}
\affiliation{School of particles and accelerators, Institute for
Research in Fundamental Sciences (IPM), P.O. Box 19395-5746,
Tehran, IRAN}
\begin{abstract}
Recently a model has been proposed that links dark matter and
neutrino masses.  The dark matter candidate which is dubbed as
SLIM has a mass of  MeV scale and can show up at low energy
experiments. The model also has  a high energy sector which
consists of a scalar doublet, $(\phi^-, \  \phi^0)$. We discuss
the  potential of the LHC for discovering the new scalars. We
focus on the $\phi^+\phi^-$ and $\phi^{\pm} \phi^0$ production and
the subsequent decay of the charged scalar to a charged lepton and
the SLIM which appears as missing energy. Identifying the
background, we estimate the signal significance and find that it
can exceed $5 \sigma$ at 30~${\rm fb}^{-1}$ for the 14 TeV run at
the LHC. We comment on the possibility of extracting the flavor
structure of the Yukawa couplings which also determine the
neutrino mass matrix. Finally, we discuss the prospects of this
search at the current 7 TeV run of the LHC.

\end{abstract}
 \maketitle
\section{Introduction}
The nature of dark matter and tiny but nonzero values of neutrino
masses are two unsolved mysteries in particle physics and
cosmology in early 21st century. Various models have been
developed that explain nonzero neutrino masses. Rich literature
also exists on the models suggesting a dark model candidate.
Attempts to link these two mysteries and to solve them within a
unique framework have only recently been started. In particular,
in Ref.~\cite{scenario} a simple low energy scenario has been
introduced which explains both phenomena in a minimalistic way.
The scenario adds a real scalar which plays the role of the dark
matter as well as two (or more) right-handed Majorana neutrinos
that along with the scalar couple to the left-handed neutrinos.
Through this Yukawa coupling, the left-handed neutrinos acquire a
small Majorana mass at one-loop level. A $Z_2$ symmetry under
which all new particles are odd is imposed to guarantee the
stability of the lightest new particle which plays the role of the
dark matter and is dubbed as SLIM. The $Z_2$ symmetry  forbids a
Dirac mass term for neutrinos. Within this scenario the main
annihilation mode for dark matter is to (anti-)neutrino pairs. To
simultaneously satisfy upper bound on the neutrino masses and to
account for the observed dark matter abundance within thermal
scenario ({\it i.e.,}  $\langle \sigma({\rm DM}+{\rm DM} \to {\rm
anything}) v \rangle \simeq 3 \cdot 10^{-26}~{\rm cm}^3/{\rm
sec}$), at least one of the right-handed neutrinos as well as the
scalar have to be lighter than 10 MeV. On the other hand, these
conditions lead to a lower bound on the couplings which makes the
scenario testable in low energy experiments such as the
measurement of ${\rm Br}(\pi , K \to l_\alpha+{\rm missing ~
energy})$ \cite{Farzan}.

With this minimal content, the new Yukawa coupling is not
invariant under the electroweak symmetry so the low energy
effective scenario has to be embedded in a
SU(3)$\times$SU(2)$\times$U(1) invariant model at higher energies.
An example of such a model is introduced in \cite{yasaman}. The
model includes a scalar electroweak doublet which has Yukawa
couplings with charged leptons and the right-handed neutrinos.
Like in the case of scenario \cite{scenario}, this Yukawa coupling
gives rise to the neutrino masses at one-loop level. The model
inherits the testability at low energy experiments from the low
energy scenario that has been embedded in it. In addition,
observable effects are expected in high energy experiments such as
the LHC as well as in searches for lepton flavor violating decays
such as $\mu \to e \gamma$. If the masses of the components of the
electroweak doublet are not too high, these scalar particles can
be produced through electroweak interactions at the LHC. As
indicated in Ref. \cite{yasaman}, it is in principle possible to
derive the flavor structure of the Yukawa couplings at the LHC and
compare it with the flavor structure deduced from the neutrino
mass matrix.

In the present paper, we elaborate on this idea in detail.
Considering various backgrounds and by employing  state-of-the-art
techniques for reconstructing the signal, the possibility of
discovering the new particles is studied. We then discuss and
propose conditions under which the Yukawa couplings can be
derived. Several packages are linked together to ensure a
reasonable framework for the simulation and the analysis of
events.

The paper is organized as follows. In sect. \ref{model}, a brief
description of the model is presented. In sect. \ref{sbs}, the
signal and background events are identified and tools for
establishing the simulation and analysis framework are introduced.
The signal and background cross sections are then calculated. In
sect. \ref{kinematics}, the number of expected events at LHC
collected with a beam of 14 TeV center of mass
 energy after 30
$fb^{-1}$ integrated luminosity  is calculated. Full set of
kinematic cuts are designed step
 by step and their efficiencies on both signal and background events
 are calculated and shown in tables. The signal significance for each
 studied category and each final state is calculated.
 In sect. \ref{parameters},
 a brief discussion
on the possibility of measuring the model parameters is presented.
In sect.~\ref{alternative}, an alternative discovery channel is
presented. In sect.~\ref{discovery}, the discovery potential of
the 7 TeV run of the LHC is discussed.
 The paper ends with a
conclusion section on the discovery potential of the model and its
new particles.

\section{The model\label{model}}
The model has a minimalistic content and adds only the following
fields to the Standard Model:
\begin{itemize}
\item  Two (or more) right-handed neutrinos, $N_i$;
\item A scalar SU(2) doublet $\Phi$ which carries hypercharge:
$\Phi=(\phi^0 \ \phi^-)$ where $\phi^0=(\phi_1+i\phi_2)/\sqrt{2}$
with real $\phi_1$ and $\phi_2$;
\item  A singlet scalar $\eta$.
\end{itemize}
A $Z_2$ symmetry is imposed on the model under which these new
particles are odd but the SM particles are even.
 The $Z_2$ symmetry forbids the
Yukawa coupling of form $\bar{N} \Phi\cdot L$  and makes the
lightest particle
 stable. The general $Z_2$ invariant
renormalizable Lagrangian involving the scalars can be written as
\begin{align}
\mathcal{L}=&-m_\Phi^2 \Phi^\dagger \cdot \Phi -{m_s^2\over 2}
\eta^2
- (m_{\eta \Phi} \eta(H^T (i\sigma_2) \Phi)+{\rm H.c.}) \nonumber \\
~&-{\lambda_1 } |H^T (i\sigma_2) \Phi|^2-{\rm Re}[\lambda_2 (H^T
(i\sigma_2) \Phi)^2]-\lambda_3 \eta^2 H^\dagger H-\lambda_4
\Phi^\dagger \cdot \Phi H^\dagger \cdot H
\nonumber \\
~& -{\lambda'_1 \over 2} (\Phi^\dagger \cdot
\Phi)^2-{\lambda'_2\over 2} \eta^4-\lambda'_3\eta^2 \Phi^\dagger
\cdot \Phi \nonumber \\
~& -m_H^2 H^\dagger\cdot H- {\lambda \over 2}(H^\dagger \cdot H)^2 \ .
\label{MainLag}
\end{align}
We are interested in the regime that only the standard Higgs, $H$,
acquires a vacuum expectation value so the $Z_2$ symmetry is
maintained. After the electroweak symmetry breaking, the third
term in Lagrangian (\ref{MainLag}) leads  to a mixing between
$\eta$ and the neutral component of $\Phi$. For simplicity, we
take the CP invariant case so $m_{\eta \Phi}$ will be real and as
a result, $\eta$ will mix only with the CP-even component of
$\phi^0$. The mass eigenstates $\delta_1$ and   $\delta_2$ can be
written as \ba \label{mixing-ETA-PHI1} \left[
\begin{matrix}\delta_1 \cr \delta_2 \end{matrix}\right]=
\left[
\begin{matrix} \cos \alpha & -\sin \alpha \cr
\sin \alpha & \cos \alpha \end{matrix} \right]\left[
\begin{matrix} \eta \cr \phi_1 \end{matrix} \right]\ea with
\begin{align}
\tan 2 \alpha &= {2 v_H m_{\eta \Phi} \over
m_{\phi_1}^2-m_\eta^2} \ ,\label{2ALPHA}\\ m^2_{\delta_1} &\simeq
m_\eta^2- {(m_{\eta
\Phi} v_H)^2 \over m_{\phi_1}^2-m_\eta^2}~~~~{\rm and}\\
m^2_{\delta_2} &\simeq m_{\phi_1}^2+ {(m_{\eta \Phi} v_H)^2 \over
m_{\phi_1}^2-m_\eta^2}\ \end{align} in which
$$ m_\eta^2=m_s^2+\lambda_3 \frac{v_H^2}{2}$$
and $$ m_{\phi_1}^2=m_{\Phi}^2+\lambda_4 {v_H^2 \over 2}+\lambda_1
{v_H^2 \over 2 }+\lambda_2{v_H^2 \over 2}.$$ The CP-odd component
of $\Phi$ remains mass eigenstate with mass eigenvalue
$$ m_{\phi_2}^2=m_{\Phi}^2+\lambda_4 {v_H^2 \over 2}+\lambda_1
{v_H^2 \over 2}-\lambda_2{v_H^2 \over 2}.$$ The mass of the
charged component of $\Phi$ is
$$m_{\phi^-}^2=m_{\Phi}^2+\lambda_4 {v_H^2 \over 2}.$$

Terms involving right-handed neutrinos are
 \be \mathcal{L}=-g_{i
\alpha}\bar{N}_{i} \Phi^\dagger\cdot L_\alpha -{M_{N_i}\over 2}
\bar{N}_i^c N_i\ \label{sterileL},\ee where  $L_\alpha$ is the
lepton doublet of flavor $\alpha$: $L_\alpha^T=(\nu_{L\alpha} \
\ell_{L\alpha}^-)$. As shown in \cite{yasaman}, this coupling at
one-loop level leads to the neutrino mass matrix as follows
\be\label{massMatrix} (m_\nu)_{\alpha \beta}=\sum_i
g_{i\alpha}g_{i\beta}A_i^2 \ee where
$$A_i^2\equiv \frac{m_{N_i}}{32\pi}\left[ \sin^2
\alpha(\frac{m_{\delta_2}^2}{m_{N_i}^2-m_{\delta_2}^2}\log
\frac{m_{N_i}^2}{m_{\delta_2}^2}-\frac{m_{\delta_1}^2}
{m_{N_i}^2-m_{\delta_1}^2}\log \frac{m_{N_i}^2}{m_{\delta_1}^2})
\right.$$
$$+\left.\frac{m_{\phi_2}^2}{m_{N_i}^2-m_{\phi_2}^2}\log
\frac{m_{N_i}^2}{m_{\phi_2}^2}-\frac{m_{\delta_2}^2}
{m_{N_i}^2-m_{\delta_2}^2}\log
\frac{m_{N_i}^2}{m_{\delta_2}^2}\right] \ .
$$
  {With
only one right-handed neutrino, the neutrino mass matrix will have
two zero mass eigenvalues and cannot accommodate the neutrino
data. The model has to include at least two right-handed
neutrinos.} From Eq.~(\ref{massMatrix}), it is straightforward to
show that with only two right-handed neutrinos, the determinant of
the neutrino mass matrix vanishes which means one of the neutrino
mass eigenvalues is zero and the neutrino mass scheme is
hierarchical. By adding another right-handed neutrino
non-hierarchical neutrino mass scheme can also be obtained. Here,
we however concentrate on the most economic case with only two
right-handed neutrinos.
 In the
appendix, we derive constraints on the flavor structure of the
couplings from the neutrino mass matrix. As shown in the appendix,
in the case of inverted hierarchical scheme $|g_{i \mu}|\simeq
|g_{i \tau}|$. Two specific solutions leading to normal hierarchical mass
schemes for neutrinos are shown in Tab. \ref{par}. To obtain these results,
we have set $\Delta m_{sun}^2=8 \times 10^{-5}~{\rm eV}^2$ and
$\Delta m_{atm}^2=2.5 \times 10^{-3}~{\rm eV}^2$  and $\lambda_2=0$ which
leads to $m_{\phi_1}=m_{\phi_2}$. $N_2$ is taken to be heavy and the
mixing between $\eta$ and $\phi_1$ is taken to be 0.01. $\theta$
 is the arbitrary mixing angle that appears in the coupling structure
(see Eqs. (\ref{GiAlpha},\ref{O})).  Notice
 that by varying $m_{\phi_1}$ and $\lambda_2$ ({\it i.e.,} parameters
 determining the heavy sector masses), the flavor structure ({\it i.e.,}
 $g_{i \alpha}/g_{i \beta}$) does not change.
Taking $m_{\phi_1}=m_{\phi_2}=150$~GeV, $g_{1 \alpha}$
 is derived from Eq.~(\ref{GiAlpha}) in the appendix and shown in the last
 line of the table.

In this model, $\delta_1$, being the lightest new particle and
therefore stable, plays the role of the dark matter. We shall call
$\delta_1$ ``SLIM''.
 Dark matter
abundance is fixed by \be \langle \sigma(\delta_1\delta_1 \to
\nu_{L\alpha} \nu_{L \beta}) v_r\rangle =\langle
\sigma(\delta_1\delta_1 \to \bar\nu_{L\alpha} \bar\nu_{L \beta})
v_r\rangle= {\sin^4 \alpha \over 8 \pi}\left| \sum_i {g_{i\alpha }
g_{i\beta } m_{N_i} \over m_{\delta_1}^2+m_{N_i}^2} \right|^2=3
\cdot 10^{-26}~{\rm cm}^3/{\rm sec} \ . \ee As shown in
\cite{yasaman}, by combining the information on $\sigma_{tot}$
from the dark matter abundance with $m_\nu$, we find
$$ m_{\delta_1}<m_{N_1}\sim {\rm few~MeV}\ .$$
That is while the other scalar particles, $\delta_2$, $\phi_2$ and
$\phi^-$ have to be heavy enough to avoid the direct bounds from
accelerators. The strongest model-independent direct bound on the
charged Higgs is 80 GeV \cite{chargedHiggs} and the bound on
$m_{{\delta}_{2}}$ and $m_{\phi_2}$ is about 90 GeV
\cite{neutralHiggs}. {At first sight, it might seem
counterintuitive that while $W^-$ and $\phi^-$ with
$m_{\phi^-}=80$~GeV have the same mass and charge, the former is
already discovered but such a $\phi^-$ has escaped detection. This
can be explained by the difference in their spin: In the process
$f \bar{f} \to \gamma^*\to \phi^-\phi^+$, the angular momentum
conservation implies that the final particles should be produced
in the $p$-wave which gives rise to a suppression by an extra
factor of $( 1-4 m_{\phi^-}^2/s)$ which for the maximum LEP energy
amounts to 0.4. Moreover, the $W^-W^+$ production can take place
via the $t$-channel neutrino exchange which has no counterpart in
the $\phi^+\phi^-$ production. To our best knowledge, no analysis
has been carried out on the Tevatron data that applies to our
model. In principle, valence quarks and antiquarks inside the
proton and antiproton beams at Tevatron can give rise to the
$\phi^+\phi^-$ production via the $s$-channel $Z^* \ , \gamma^*$
exchanges. However, each of the valence quarks on average carries
only about 0.16 of the proton energy so the effective center of
mass energy will be around 300 GeV($\simeq0.16\times 2$ TeV). As a
result like the LEP case, the production rate at Tevatron will be
suppressed.
 The same process also plays the dominant role in the
$\phi^+\phi^-$ production at the LHC  but in the case of the LHC,
the antiquark taking part in the production will be a sea quark
which on average carries a lower fraction of the proton energy,
$\langle x_{\bar q}\rangle \simeq 0.027$.
 The average center of mass energy of quark antiquark pair for the
 14 TeV run will be $
 \sqrt{\langle x_q \rangle \langle  x_{\bar q}\rangle}
 E_{cm}\simeq 800$ GeV
  which is still higher than that for the Tevatron
 despite the fact that in the case of the Tevatron, the quark
 antiquark pair are both valence quarks but in the case of the LHC
 one of the quarks is a sea quark.
We shall discuss the effect in detail. As we shall demonstrate the
discovery chance rapidly increases with center of mass energy so
the discovery potential of the Tevatron cannot be significant
comparing to that of the LHC. In addition to quark antiquark
annihilation mode, $\phi^+\phi^-$ can be produced via gluon
fusion.
  Similar consideration
holds for the production of $\delta_2$ and $\phi_2$. As we shall
see, the suppression discussed above
 has the negative consequence
for our analysis: The signal for the $\phi^-$ production suffers
from large background from the $W^-$ production.}

This model has similarities with the so-called inert model
\cite{inert} but here we have an extra singlet scalar and the main
annihilation mode of dark matter pair is into neutrinos.
Nevertheless, the contribution to the oblique parameters in our
model is similar to that in the inert model. Like the inert model,
the new contribution to the oblique parameters can cancel the one
from a SM Higgs so heavier masses for Higgs can be made compatible
with the SM. Throughout this paper, we assume that $m_{\phi^-}$ is
equal or lighter than $m_{\delta_2}$ and $m_{\phi_2}$. Moreover,
we shall assume that $m_{\phi_2}-m_{\phi^-}$,
$m_{\delta_2}-m_{\phi^-}$ and $|m_{\delta_2}-m_{\phi_2}|$ do not
exceed 80 GeV to forbid two body decays $\delta_2,\phi_2 \to
W^+\phi^-$ or $\phi_2 \to \delta_2 Z $ (or $\delta_2 \to \phi_2 Z
$). This assumption is just for simplicity of the analysis.
Probing the whole parameter space is beyond the scope of the
present paper.
\begin{table}
\begin{center}
\begin{tabular}{|l|c|c|}
\hline
& Point A & Point B \\
\hline
$m_{N_{1}}$~({\rm MeV}) & 1 & 1 \\
\hline
$m_{N_{2}}$ & $> m_{\phi^-}$ & $>m_{\phi^-}$ \\
\hline
$\alpha$ & 0.01 & 0.01 \\
\hline
$\lambda_{2}$ & 0 & 0 \\
%
$m_{\phi_2}$~({\rm GeV}) & 90 & 90 \\
\hline
$\theta$ & $\pi / 2$ & 0 \\
\hline\hline
$g_{1\alpha }$ & $\left( \begin{array}{c} 0 \\ 0.03 \\ 0.03 \end{array} \right)$ & $\left( \begin{array}{c} 0.01 \\ 0.01 \\ -0.01 \end{array} \right)$ \\
\hline
\end{tabular}
\end{center}
\caption{Model parameters.\label{par}}
\end{table}

\begin{figure}
\begin{center}
\includegraphics[width=0.8\textwidth]{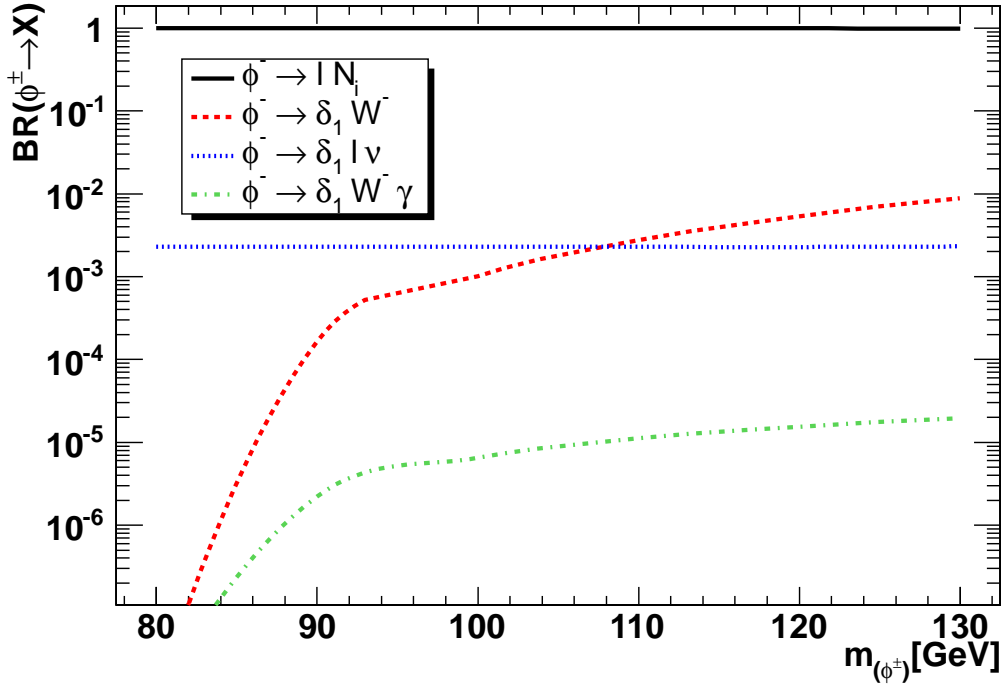}
\end{center}
 \caption{\label{br}Branching ratio of $\phi^{\pm}$
decays as a function of its mass. Point A from Tab. \ref{par}
has been used as input for this calculation.}

\end{figure}

In this model, $\phi^-$ dominantly decays via the $g_{i \alpha}$
couplings: \be \Gamma(\phi^- \to l_\alpha N_i)  =\frac{|g_{i
\alpha}|^2}{ 16 \pi}
\frac{(m_{\phi^-}^2-m_{N_i}^2)^2}{m_{\phi^-}^3}~~~~~{\rm for} ~~
m_{N_i}<m_{\phi^-} \ . \ee As discussed earlier, $N_1$ has to be
light so decay modes $\phi^- \to N_1 e^-$, $N_1\mu^-$ and
$N_1\tau^-$ are guaranteed to exist. If the mass of the other
right-handed neutrino, $N_k$, is lighter than $m_{\phi^-}$ but
$m_{N_k}\sim m_{\phi^-}$,  by studying the energy spectrum of the
emitted charged lepton, one can in principle distinguish between
this decay and $\phi^- \to \ell_\alpha^- N_1$. Moreover, $M_{N_k}$
and $|g_{k \alpha}|$ can in principle be derived. If for all $k$
larger than one $m_{N_k}>m_{\phi^-}$, the decay mode $\phi^-\to
N_k \ell_\alpha^-$ is not open so the decay mode ($\phi^- \to
\ell_\alpha^-$+missing energy) gives only $|g_{1 \alpha}|^2$. In
case that $m_{N_k}\sim m_{N_1}$, it will not in practice  be
possible to distinguish between $N_1$ and $N_k$ so the signal
($\phi^- \to \ell_\alpha^-+$ missing energy) will practically
yield $\sum_i|g_{i \alpha}|^2$ where the sum runs over all $N_i$
for which $m_{N_i}<m_{\phi^-}$. In this paper, we focus on the
case with only two $N_i$ with $m_{N_2}>m_{\phi^-}$. Apart from
this decay mode, $\phi^-$ can have other decay modes such as
$\phi^- \to W^- \delta_1$, $\phi^-\to \delta_1 \ell_\alpha^- \nu$
and $\phi^-\to \delta_1 W^- \gamma$ which as seen from
Fig.~\ref{br} are all subdominant for values of $g_{1
\alpha}\stackrel{>}{\sim} 0.01$ and $\sin \alpha\stackrel{<}{\sim}
0.01$.

The hypercharge of the new doublet, $\Phi$, is the same as that of
$H_d$ in the MSSM. Thus, in the limit that $\langle H_d \rangle$
vanishes ({\it i.e.,} $\tan \beta \to 0$) and the $\eta$-$\phi$
mixing goes to zero ({\it i.e.,} $\alpha \to 0$), the gauge
interactions of $\phi_2$ and $\phi^\pm$ are similar respectively
to those of $A^0$ and $H^\pm$. Depending on the value of
$m_{\phi_1}$, $\delta_2$ will have similar gauge interactions as
any of the CP-even Higgses, $h$ or $H$ within the MSSM. The pair
production of these particles at the LHC takes place through the
electroweak interactions. That is the $\phi^+\phi^-$, $\phi^\pm
\phi_2$ and $\phi^\pm \delta_2$ production are respectively
similar to $H^+H^-$, $H^\pm A^0$ and $H^\pm h (H)$ production
within the MSSM with $\tan \beta \to \infty$. However, the Yukawa
couplings of $\Phi$ in our model is quite different from those of
$H_d$ within the MSSM. As a result, the decay processes for the
$\Phi$ components differ from those for the Higgs components.
Neutral CP-odd Higgs within the MSSM can be singly produced via
gluon fusion but the parallel does not exist within this model in
which Yukawa coupling between $\Phi$ and the $b$ quarks are
forbidden.

To perform our analysis, we focus on the points shown in Tab.~\ref{par}.
The point A in Table \ref{par} is chosen to search for
 tauonic and muonic final
states while the point B is used to search for final states
involving electrons. The point A in the parameter space leads to a
negligible branching ratio for $\phi^{-}$ decay to electrons while
decay to a $\mu$ or $\tau$ lepton is possible with equal
probability. That is at point A,
$\textnormal{BR}(\phi^{\pm}\rightarrow \mu^\pm
N_1)=\textnormal{BR}(\phi^{\pm}\rightarrow \tau^\pm N_1) \simeq
0.5$. Figure \ref{br} shows the contributions from all the
 decay
channels and verifies that the above approximation is valid up to
a maximum $1\%$ error
 for $m_{(\phi^{\pm})}=130$ GeV.\\
In this scenario three samples of events are expected
corresponding to different final states:
$\mu\mu$, $\mu\tau$ and $\tau\tau$.
The $\tau\tau$ final state suffers
 from the large hadronic backgrounds such
 as $W+$jets with $W$ decaying to $\tau$ or jets.
  These background events may survive due to  light jets faking the $\tau$.\\
As we shall see, the $\mu\mu$ and $\mu\tau$ final states can be observed
 with a significance
exceeding $5\sigma$ at $30 fb^{-1}$.
The next sections are devoted to signal and background simulation.
The event analysis and selection is then described based on
 the kinematic cuts applied in order to increase the signal to
  background ratio. A full study of kinematic
  distributions is performed and a dedicated $\tau$
  jet identification is applied similar to
  the standard LHC algorithms in Ref. \cite{tauid}.
  Finally the signal significance is estimated for each search
category and final state.

\section{Signal and Background Simulation \label{sbs}}
 Considering the above observation the $\phi^{\pm}$
pair production  is generated using the process
 $pp \rightarrow H^{+}H^{-}$ setting $\tan\beta=100$
 which is reasonably high.
These events are generated using PYTHIA 8.1.2.5  \cite{pythia}. To
account for the $\tau$ leptons proper decays, TAUOLA package  \cite{tauola} is used.
 Since PYTHIA 8 is based on C++ programming language,
 TAUOLA C++ interface  \cite{tauola_interface} is used to link
 TAUOLA and PYTHIA 8 for the correct production of
 $\tau$ hadronic decays. Since TAUOLA interface output is in the
 HepMC format, HepMC 2.5.1  \cite{hepmc}
 is linked to TAUOLA interface and is used in the analysis.\\
Throughout the analysis, the $\tau$ leptons are identified
through their hadronic decays which results in a narrow $\tau$ jet which will be described in detail later in the next sections.\\
The jet reconstruction is performed using FASTJET 2.4.1
\cite{fastjet} with anti-kt algorithm  \cite{antikt} and a cone
size of $0.4$ and the $E_{T}$ recombination scheme.

\begin{figure}
\begin{center}
\includegraphics[width=0.80\textwidth]{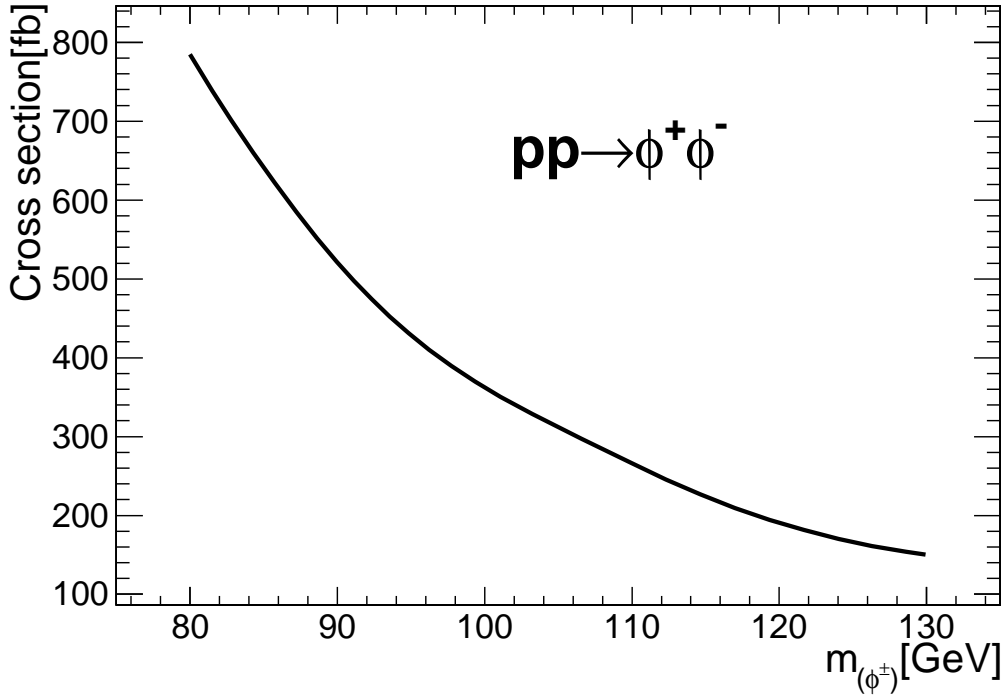}
\end{center}
\caption{Signal cross section as a function of $m_{\phi^{\pm}}$.\label{Xsec}}
\end{figure}

The signal cross section is calculated using PYTHIA. As mentioned
earlier, the charged Higgs pair production is expected to simulate
$\phi^{\pm}$ events if a large enough $\tan\beta$ is used. Since
the signal cross section decreases rapidly with increasing
$m_{\phi^{\pm}}$, the low mass $\phi^{\pm}$ is studied in this
analysis. Figure \ref{Xsec} shows the signal cross section as a
function of $m_{\phi^{\pm}}$.
The main background events are $W^{+}W^{-}$, $t\bar{t}$, $W+$jets
and $Z+$jets. Their cross sections are calculated using MCFM 5.7
\cite{mcfm} at NLO.
Table \ref{bXsec} shows the background cross sections calculated
using MCFM with renormalization and factorization scale set to the
top quark mass (172.5 GeV) for $t\bar{t}$ events, the $W$ boson mass
(80.4 GeV) for $W^{+}W^{-}$ and $W+$jets events and the $Z$ boson mass
(91 GeV) for $Z+$jets events. For both event simulation and cross
section calculation purposes, MRST 2004 NNLO PDF is used by linking
LHAPDF 5.8.1  \cite{lhapdf} to the PYTHIA event generator as well as
MCFM. \\
In order to calculate the expected number of events at
30$fb^{-1}$, the total cross sections in Table \ref{bXsec} are
multiplied by the proper branching ratios. For $t\bar{t}$ events,
while one of $W$ bosons (from $t$ or $\bar{t}$ decay) decays to a
muon, the other $W$, is forced to decay to a $\tau$ lepton or
light jets. The same decay channel is considered for $W^+W^-$
events. The $W$ boson decay to light jets is considered to account
for the $\tau$ jet fake rate from light jets. For $W$+jet events,
the $W$ boson is forced to decay to a muon and appearance of a
$\tau$ jet is only possible again through the $\tau$ jet fake
rate. The $Z+$jets process is only considered for the $\mu\mu$
final state and the $Z$ boson is forced to decay to a pair of
muons. In this analysis the following values are used for
branching ratio of $W$ and $Z$ bosons according to Particle Data
Group  \cite{pdg}: BR$(W^+ \rightarrow \mu^+\nu)=10.57\%$, BR$(W^+
\rightarrow \tau^+\nu)=11.25\%$, BR$(W^+ \rightarrow
\textnormal{hadrons})=67.60\%$, BR$(Z \rightarrow \mu\mu)=3.36\%$.

\begin{table}
\begin{center}
\begin{tabular}{|l|c|c|c|c|}
\hline
Process & $W^{+}W^{-}$ & $t\bar{t}$ & W+jets & Z+jets \\
\hline
Cross Section & 115.5$\pm$0.4 pb& 878.7$\pm$0.5 pb& 187.1$\pm$0.1 nb & 258.9$\pm$0.7 nb \\
\hline
\end{tabular}
\end{center}
\caption{Background cross sections calculated using MCFM package.\label{bXsec}}
\end{table}
As is seen from Table \ref{bXsec}, a large suppression factor is
needed due to the large background cross sections. In the next
sections, it is shown that a reasonable background suppression can
be achieved by applying kinematic cuts and using a
state-of-the-art $\tau$ identification and selection.

\section{Event Kinematics\label{kinematics}}
As mentioned earlier in this analysis only two final states of
$\mu\mu$ and $\tau\mu$ from $\phi^+\phi^-$ decay are considered.
In the following,
 two separate analyses are designed and described in detail.
 It is convenient to define the transverse energy, $E_T$, as
$$ E_T \equiv \sqrt{p_x^2+p_y^2}$$ where the $z$ direction is
taken parallel to the beam direction.

\subsection{The $\tau ~ \mu ~ E^{miss}_{T}$ Final State}
\begin{figure}
\begin{center}
\includegraphics[width=0.80\textwidth]{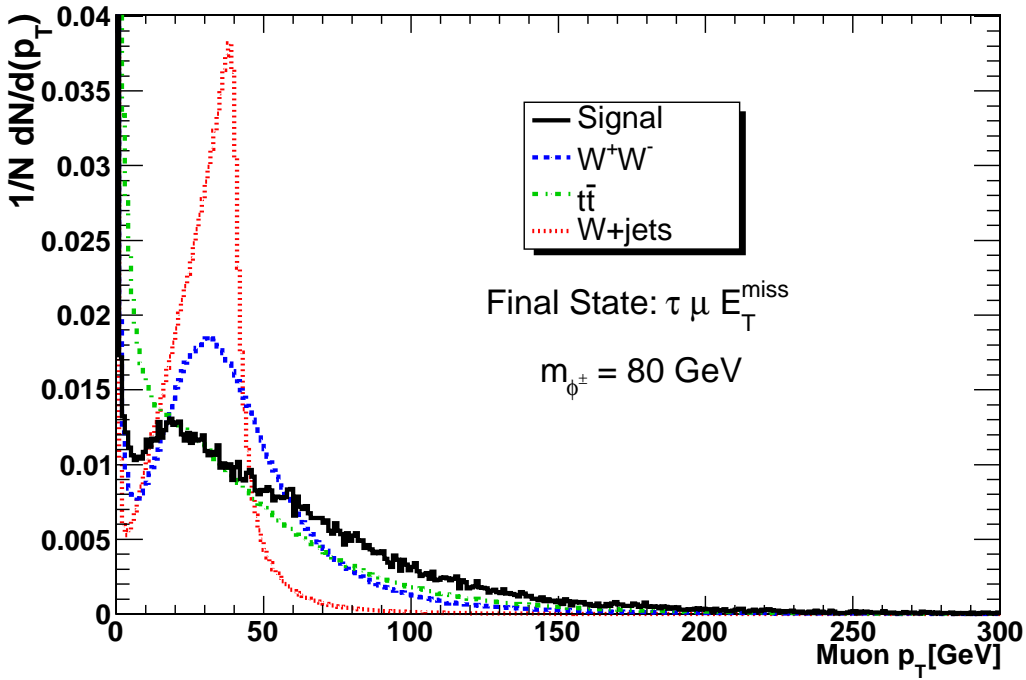}
\end{center}
\caption{Muon transverse momentum distributions in signal and background events in the $\tau\mu E^{miss}_{T}$ final state.\label{mupt}}
\end{figure}

\begin{figure}
\begin{center}
\includegraphics[width=0.80\textwidth]{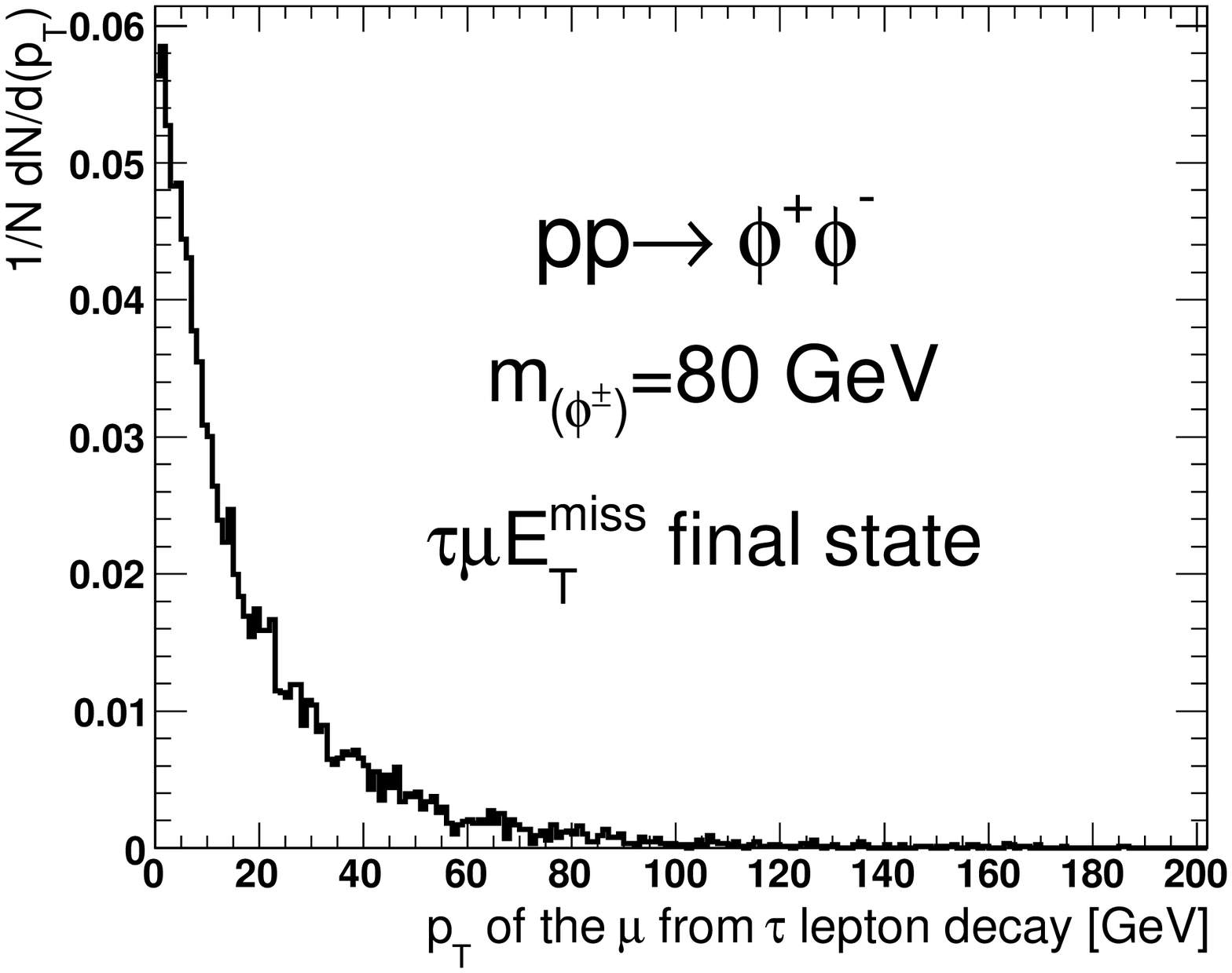}
\end{center}
\caption{Transverse momentum distribution of the muon from $\tau$
lepton decay in signal events in the $\tau\mu E^{miss}_{T}$ final state.
\label{mutau}}
\end{figure}

This final state is searched for by requiring one muon and one
$\tau$ jet. The muon transverse momentum distribution has been
shown in Fig. \ref{mupt}. The small contribution from soft muons
in $\tau$ lepton decays (see Fig. \ref{mutau}), can be viewed as a
small excess in signal and background events. In the signal
events, in the search for $\tau ~ \mu ~ E^{miss}_{T}$ final state,
this contribution may come from $pp\rightarrow \phi^{+}\phi^{-}
\rightarrow \tau\tau E^{miss}_{T}$, with one $\tau$ decaying
hadronically and the other decaying to a muon.
 However, due to the small branching ratio of $\tau$ decay to muons
 ($\sim 17\%$) and the soft $p_{T}$ distribution of these muons,
 no sizable contribution from these events is expected.
 The same considerations hold for muonic decays of the $\tau$ leptons
 in background events like $t\bar{t}$ and $WW$. As a result, the $\tau$
 leptons mainly contribute in the analysis through their
 hadronic decays. The following kinematic requirement is applied
 on muons in the event:
\begin{equation}
p^{\mu}_{T}>50 ~\textnormal{GeV} ,~ |\eta|<2.5 \ .
\label{mukin1}
\end{equation}
where $\eta$ is the pseudorapidity defined as
$$\eta\equiv -\log\left[ \tan \frac{\theta}{2} \right]$$
in which $\theta$ is the angle between the momentum and the beam axis.
This kinematic cut is applied mainly to suppress backgrounds with soft muons in the final state such as $W+$jets. \\

\begin{figure}
\begin{center}
\includegraphics[width=0.80\textwidth]{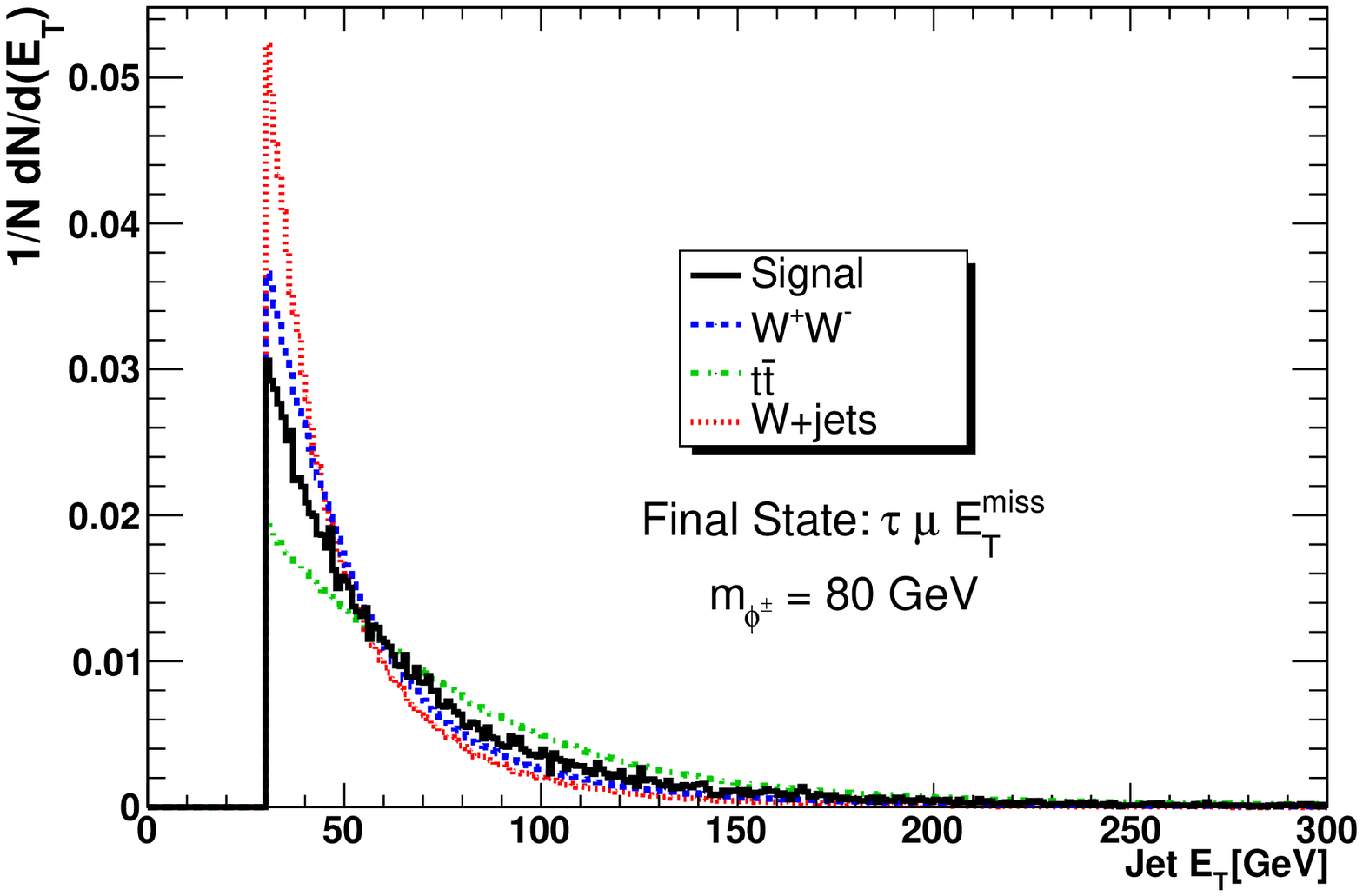}
\end{center}
\caption{Transverse energy distribution of
reconstructed jets in signal and background events in the $\tau\mu E^{miss}_{T}$ final state.
 All jets satisfying Eqs. \ref{jetetEq} and
\ref{isol} fill the histogram. Softer jets with $E_T<30$~GeV are
discarded as the jet multiplicity strongly depends on the jet
transverse energy and rapidly increases when soft jets are
included, leading to  long analysis time per event.\label{jetet}}
\end{figure}

\begin{figure}
\begin{center}
\includegraphics[width=0.80\textwidth]{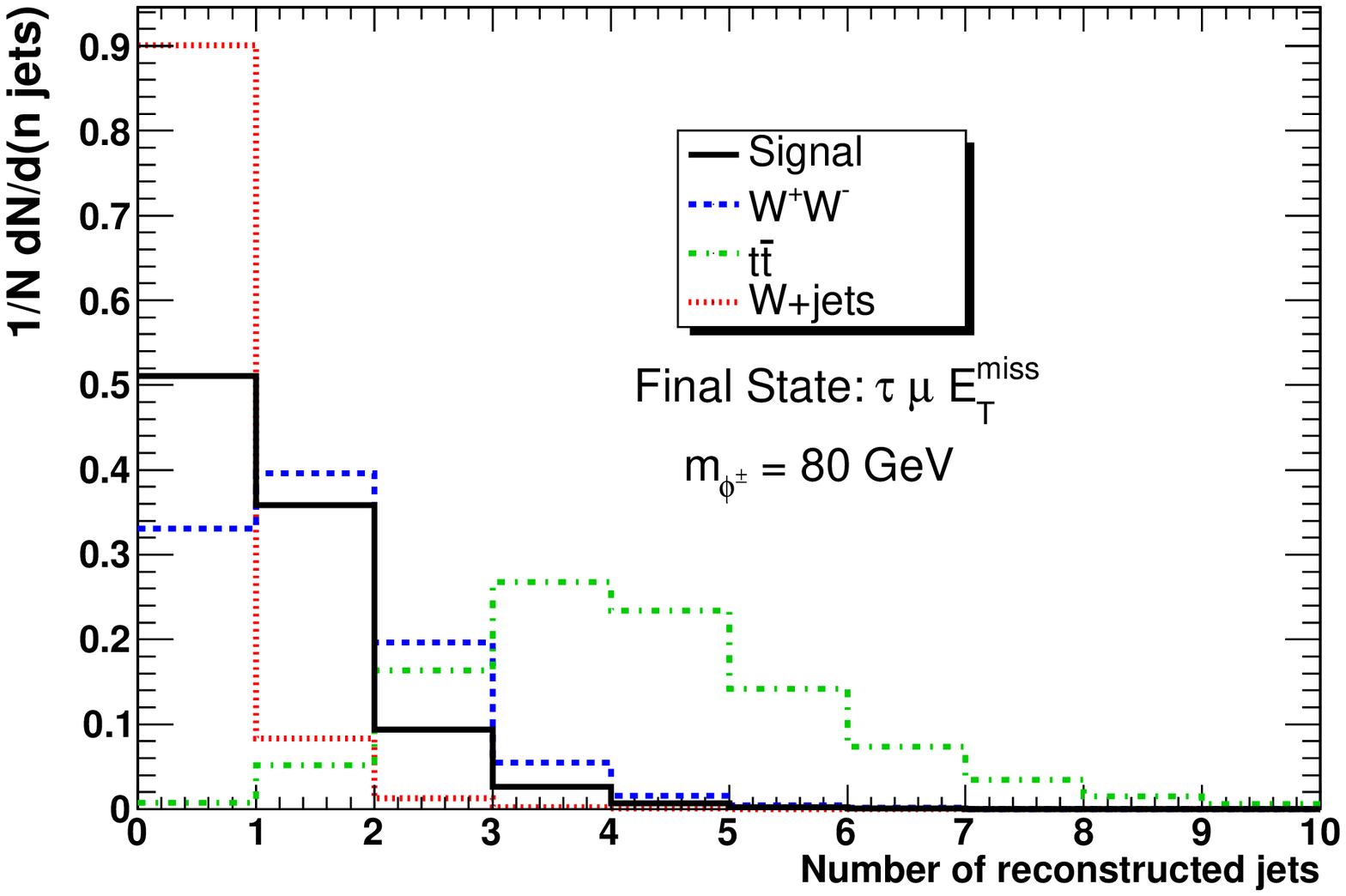}
\end{center}
\caption{Number of reconstructed jets in signal and background events
in the $\tau\mu E^{miss}_{T}$ final state. A jet is considered and
counted if it satisfies Eqs. \ref{jetetEq} and \ref{isol}.\label{njets}}
\end{figure}

\begin{figure}
\begin{center}
\includegraphics[width=0.80\textwidth]{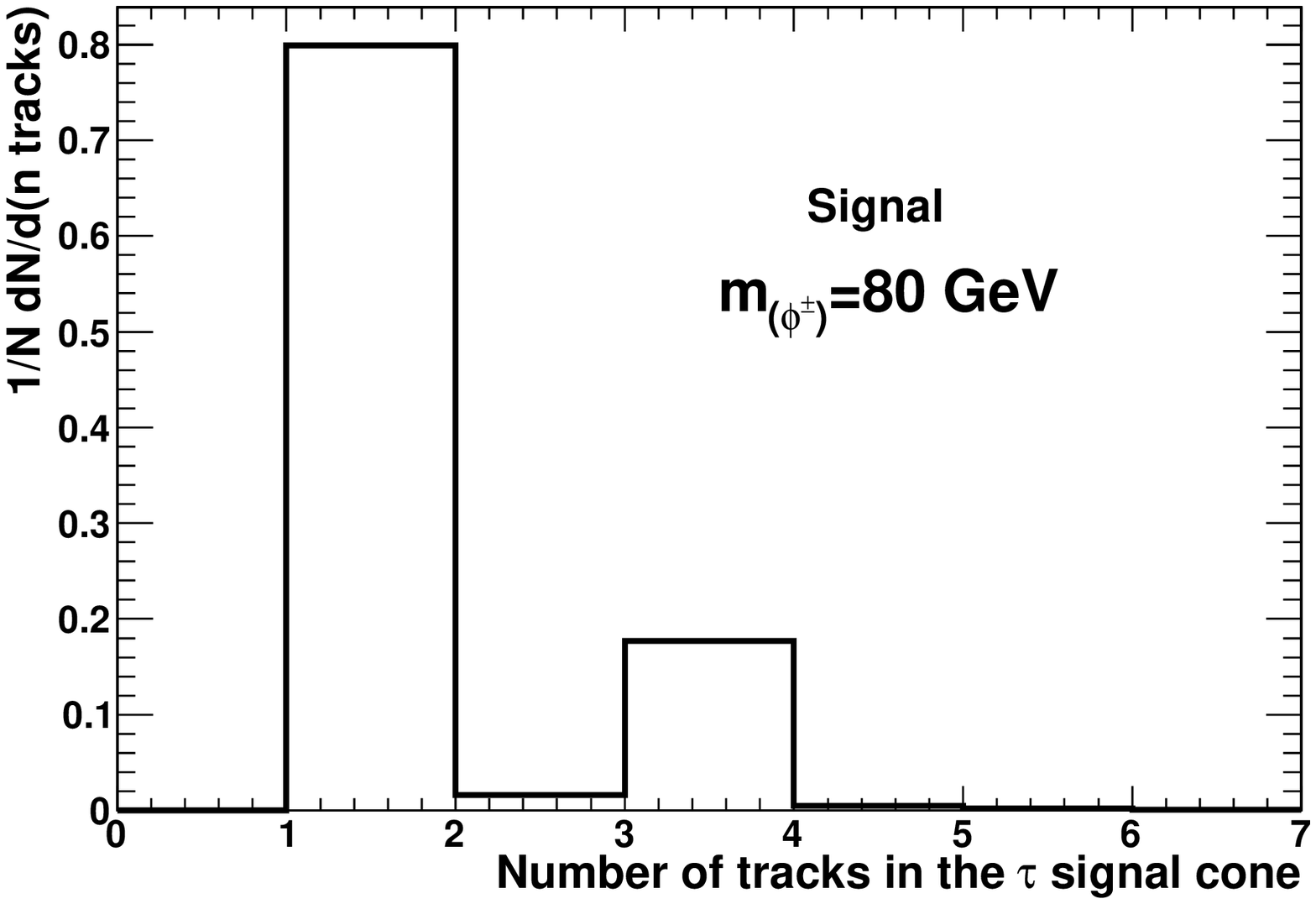}
\end{center}
\caption{Number of tracks in the signal cone around the leading
track of the $\tau$ jet.\label{nst}}
\end{figure}

\begin{figure}
\begin{center}
\includegraphics[width=0.80\textwidth]{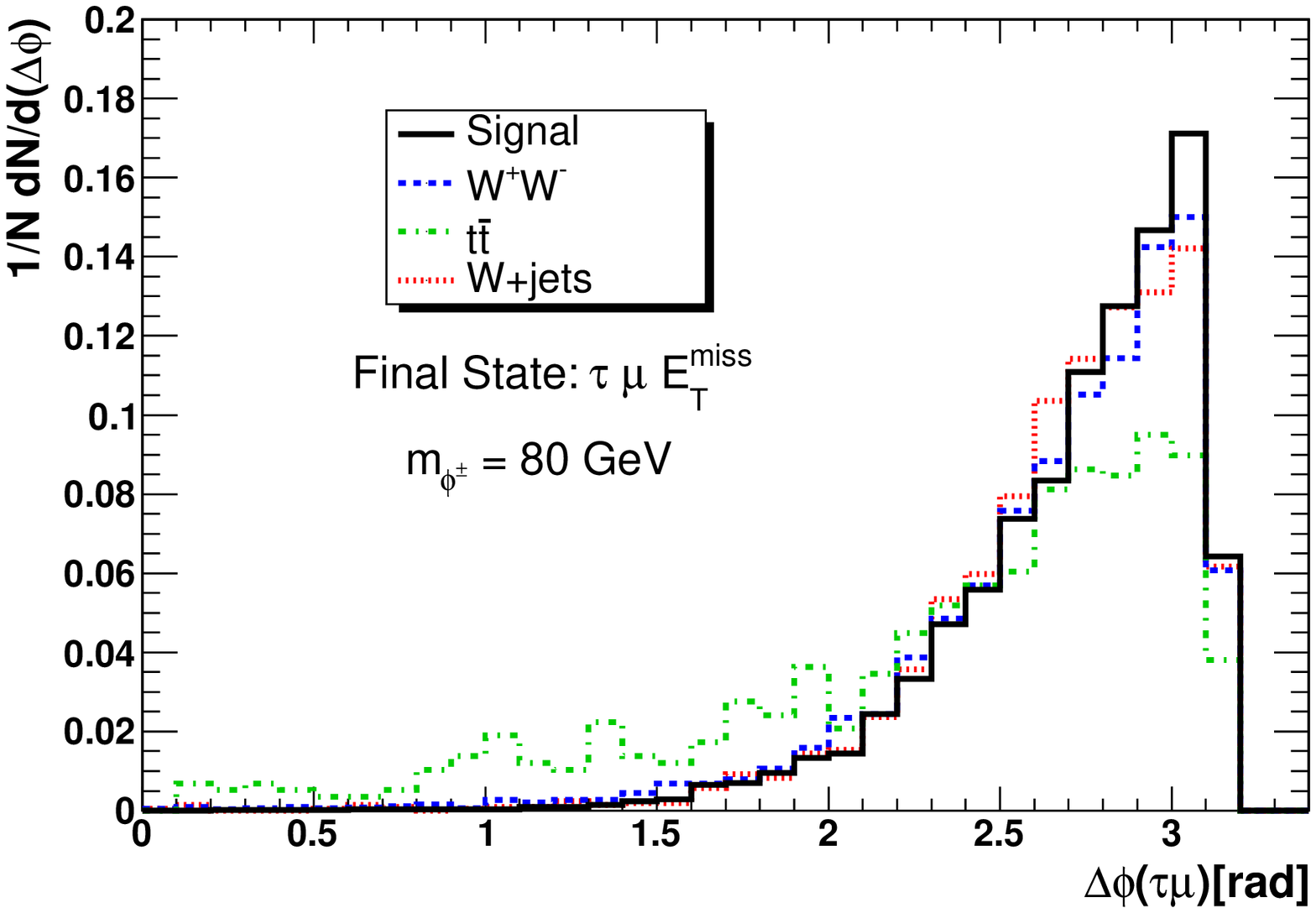}
\end{center}
\caption{The distribution of the azimuthal angle between the muon
and the $\tau$ jet candidate.\label{dphi}}
\end{figure}

The $\tau$ jet identification is performed by first reconstructing jets and requiring kinematic cuts as the following:
\begin{equation}
 E^{\textnormal{jet}}_{T}>30 ~ \textnormal{GeV}, ~ |\eta|<2.5 \ .
\label{jetetEq}
\end{equation}
A reconstructed jet which passes the above requirement is considered
as a $\tau$ jet candidate. All selected jets should be separated
enough from muons with $p_{T}>20$ GeV with the following requirement:
\begin{equation}
\Delta R_{(\textnormal{jet,$\mu$})}>0.4 \ .
\label{isol}
\end{equation}
Here $\Delta R$ is defined as $\Delta R= \sqrt{(\Delta \eta)^2+
(\Delta \phi)^2}$
where $\eta$ is the pseudorapidity and $\phi$ is the azimuthal
angle with respect to the collider beam axis. Figure \ref{jetet}
shows the transverse energy distribution of selected
jets and the number of reconstructed jets passing
above requirements are shown in Figure \ref{njets}.
 Due to the requirements of Eq. \ref{jetetEq} and \ref{isol},
  some signal events are killed, however, a larger fraction
  of background events fall in zero jet bin.
  The single $W$ production has the largest cross section
  when produced with no jet so a large fraction
  of them fill the zero jet bin. Another reason is the fact that the
  jet accompanying the $W$
   boson is usually a spectator quark flying in
  the forward-backward direction and as a result, failing
  the requirement of Eq. \ref{jetetEq}.
The $\tau$ hadronic decay produces predominantly one or three charged
pions with a sizably high transverse momentum due to the low charged
track multiplicity compared to the light quark jets.
The leading track associated to the jet is found by searching
for the highest-$p_{T}$ track in the matching cone defined around
the jet axis with an opening angle of $\Delta R<0.1$. \\
The desired polarization of the $\tau$ leptons in $\phi^{\pm}$
decays is not simulated by using charged Higgs events in PYTHIA
because of opposite polarizations between the PYTHIA charged Higgs
events and
our model. Thus, in this analysis polarization dependent cuts on leading track $p_{T}$, the ratio of leading track $p_{T}$ over $\tau$ jet energy, and $E^{miss}_{T}$ are removed. The TAUOLA package is still needed to get correct branching ratio of $\tau$ lepton decays and produce hadronic decays with proper charged track multiplicity and decay topology. The rest of $\tau$ identification algorithm is applied as the following. The isolation cone is defined with a cone size of $\Delta R<0.4$ around the leading track. The signal cone is also defined around the leading track with $\Delta R<0.07$. The isolation is applied by requiring no charged track with $p_{T}>1$ GeV to be in the isolation annulus defined as $0.07<\Delta R<0.4$. This requirement suppresses light quark jets which contain many charged tracks in their isolation annulus.\\
To further suppress the fake $\tau$ jet rate and increase the
 $\tau$ jet purity in events with genuine $\tau$ jets, a search for charged tracks in the signal cone is made and the number of tracks in the signal cone is required to satisfy the following requirement:
\begin{equation}
\textnormal{Number of signal tracks} = 1 ~ \textnormal{or} ~ 3 \ .
\end{equation}
Figure \ref{nst} shows distribution of the number of signal
 tracks in signal events before applying the cut.
  As seen from Fig. \ref{nst}, $\tau$ jets have undergone
  1- or 3-prong decays predominantly.
The muons and $\tau$ leptons are expected to
be produced back-to-back in the signal events.
 Figure \ref{dphi} shows the distribution of the
 azimuthal angle between the two leptons.
 In the case of the $\tau$ leptons, the $\tau$ jet direction is used.
As is observed from Fig. \ref{dphi}, background events also show
similar distributions. Thus, no cut on $\Delta \phi$ between the
muon and the $\tau$ jet is applied. In the next step, the $\tau$
lepton charge is calculated as the sum of charges of tracks in the
signal cone. Since muons and $\tau$ leptons in signal events are
produced with opposite charges, the following requirement is
applied:
\begin{equation}
\textnormal{Muon~ charge +~} \tau \textnormal{~jet~ charge} =0\ .
\end{equation}
The analysis is therefore  only sensitive to the sum of charges of
the two leptons and does not care whether $\mu$ and $\tau$ are
individually positive or negative.

 An event is selected if it passes all the above
selection requirements. Table \ref{sigsel} shows selection
efficiencies for signal events with $m_{(\phi^{\pm})}$ ranging
from 80 GeV to 130 GeV. The quoted efficiencies show fraction of
events which pass a cut and contain exactly the same number of
physical objects as the cut requires, {\it e.g.,} the fourth row
shows the fraction of events having exactly one muon passing the
kinematic cuts  in Eq.~(\ref{mukin1}).
\begin{table}
\begin{center}
\begin{tabular}{|l|c|c|c|c|}
\hline
$m_{(\phi^{\pm})}$ & 80 GeV & 90 GeV & 110 GeV & 130 GeV  \\
\hline
Total cross section [fb] & 783 & 521 & 265 & 150 \\
\hline
Number of events at 30 $fb^{-1}$ & 11745 & 7815 & 3975 & 2250  \\
\hline
N Muons = 1 & 5018(42.7$\%$) & 3810(48.7$\%$) & 2251(56.6$\%$) & 1419(63$\%$)\\
\hline
N Jets = 1 & 2029(40.4$\%$) & 1585(41.6$\%$) & 976(43.4$\%$) & 647(45.6$\%$)\\
\hline
Isolation & 1375(67.7$\%$) & 1101(69.4$\%$) & 686(70.2$\%$) & 468(72.4$\%$)\\
\hline
1- or 3-prong decay & 1306(95$\%$) & 1045(94.9$\%$) & 653(95.3$\%$) & 447(95.5$\%$) \\
\hline
opposite charge & 1304(99.8$\%$) & 1043(99.8$\%$) & 653(99.9$\%$) & 447(99.9$\%$)\\
\hline
total efficiency & 11.1$\%$ & 13.3$\%$ & 16.4$\%$ & 19.8$\%$ \\
\hline
Expected events at $30fb^{-1}$ & 1304 & 1043 & 653 & 447  \\
\hline
\end{tabular}
\end{center}
\caption{Selection efficiencies and remaining number of signal
events after each cut in the $\tau\mu E^{miss}_{T}$ final state.
Numbers in parentheses are relative efficiencies in percent with
respect to the previous cut. Branching ratios have been taken into
account in transition from the second to third row, thus the
number of events at 30$fb^{-1}$ has been calculated as
$N=\sigma_{tot.}\times BR \times Lumi.$, where $\sigma_{tot.}$ is
the value quoted in the second row, $BR$ is the product of all
branching ratios of decays taking into account the permutation
factors, e.g. in this analysis it is $2\times
BR(\phi^{+}\rightarrow \tau E^{miss}_{T})\times
BR(\phi^{-}\rightarrow \mu E^{miss}_{T})$, and $Lumi.$ is the
integrated luminosity which is 30$fb^{-1}$ in this
analysis.\label{sigsel}}
\end{table}

In this analysis, selection cuts are mass independent. As is seen,
the total selection efficiency increases due to harder kinematic
distributions with increasing $m_{(\phi^{\pm})}$. This effect is,
however, compensated by decreasing cross section of signal events.
Table \ref{bsel} shows selection efficiencies for background
samples.

\begin{table}
\begin{center}
\begin{tabular}{|l|c|c|c|}
\hline
Process & $W^{+}W^{-}$ & $t\bar{t}$ & W+jets \\
\hline
Total cross section [pb] & 115.5 & 878.7 & 187.1$\times 10^3$ \\
\hline
Number of events at 30 $fb^{-1}$ & 577577 & 4394086 & 5.9$\times10^8$  \\
\hline
N Muons = 1 & 142084(24.6$\%$) & 1.8e+06(40.9$\%$) & 2.5e+07(4.2$\%$)\\
\hline
N Jets = 1 &  60812(42.8$\%$) & 81512(4.5$\%$) & 1.1e+07(44.5$\%$) \\
\hline
Isolation & 8483(13.9$\%$) & 5533(6.8$\%$) & 953682(8.6$\%$)\\
\hline
1- or 3-prong decay & 4807(56.7$\%$) & 2545(46$\%$) & 244810(25.7$\%$)\\
\hline
Opposite charge & 4618(96.1$\%$) & 2241(88.1$\%$) & 209655(85.6$\%$)\\
\hline
total efficiency & 0.8$\%$ & 0.05$\%$ & 0.03$\%$ \\
\hline
Expected events at $30fb^{-1}$ & 4618 & 2241 & 209655  \\
\hline
\end{tabular}
\end{center}
\caption{Selection efficiencies and remaining number of background
events after each cut in the $\tau\mu E^{miss}_{T}$ final state.
Numbers in parentheses are relative efficiencies in percent
with respect to the previous cut. Branching ratios have been taken
 into account in transition from the second to third row.\label{bsel}}
\end{table}

Having obtained the number of signal and background events after
all selection cuts, the signal significance is now calculated as
\begin{equation}
\textnormal{Signal~ Significance}=\frac{N_{S}}{\sqrt{N_{B}}}
\label{signif}
\end{equation}
where $N_{S}~(N_{B})$ is the final number of signal (total background) events after all selection cuts.
Table \ref{sig} shows the signal significance for different $m_{(\phi^{\pm})}$ values at $30~fb^{-1}$.
\begin{table}
\begin{center}
\begin{tabular}{|l|c|c|c|c|}
\hline
$m_{(\phi^{\pm})}$ & 80 GeV & 90 GeV & 110 GeV & 130 GeV \\
\hline
Signal significance & 2.8 & 2.2 & 1.4 & 1 \\
\hline
\end{tabular}
\end{center}
\caption{Signal significance in the $\tau\mu E^{miss}_{T}$ final
state for different $m_{(\phi^{\pm})}$ hypotheses at
$30~fb^{-1}$.\label{sig}}
\end{table}

\subsection{The $\mu ~ \mu ~ E^{miss}_{T}$ Final State}
\begin{figure}
\begin{center}
\includegraphics[width=0.80\textwidth]{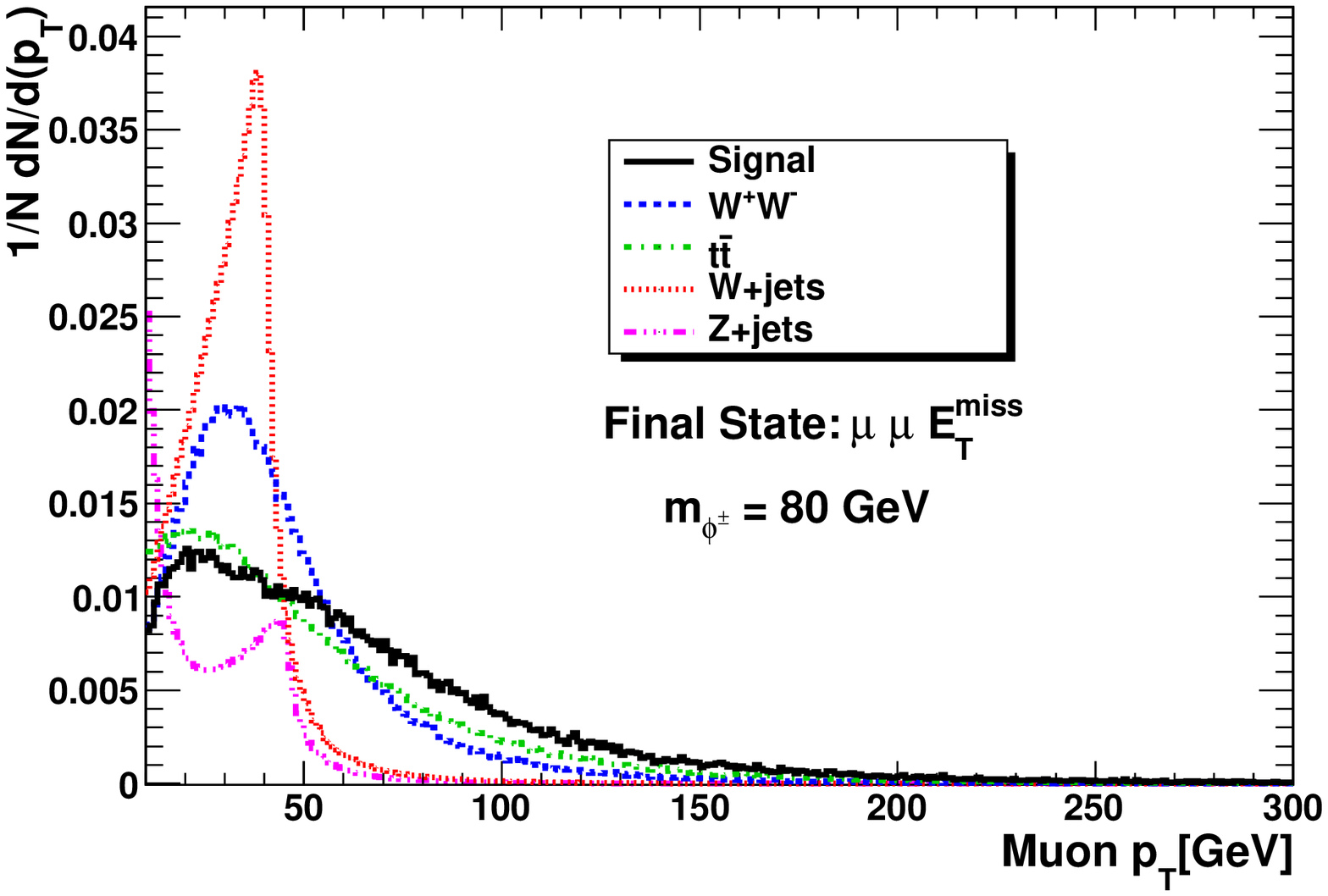}
\end{center}
\caption{Muon transverse momentum distributions in signal and background events in the $\mu\mu E^{miss}_{T}$ final state. This is an inclusive histogram i.e. any muon in the event satisfying kinematic cuts as in Eq. \ref{mukin2} has filled the histogram. Histogram starts from 10 GeV on the X-axis to suppress the large peak of soft muons from off-shell Z+jets events. \label{mupt2}}
\end{figure}

\begin{figure}
\begin{center}
\includegraphics[width=0.80\textwidth]{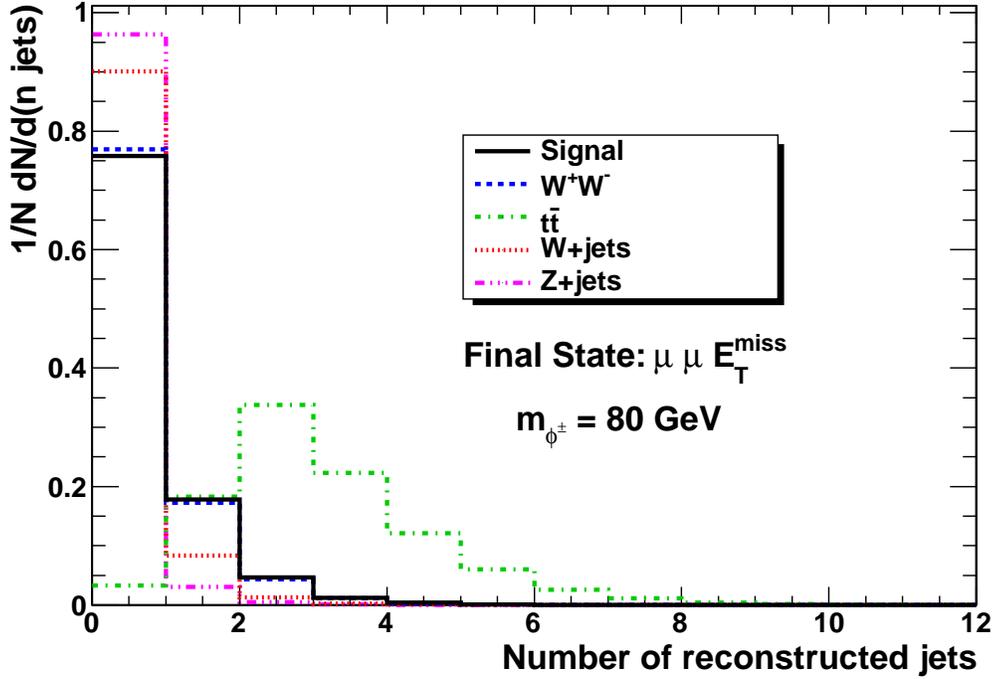}
\end{center}
\caption{Number of reconstructed jets in signal and background events in the $\mu\mu E^{miss}_{T}$ final state. A jet is considered and counted if it passes kinematic requirements described in the text.\label{njets2}}
\end{figure}

\begin{figure}
\begin{center}
\includegraphics[width=0.80\textwidth]{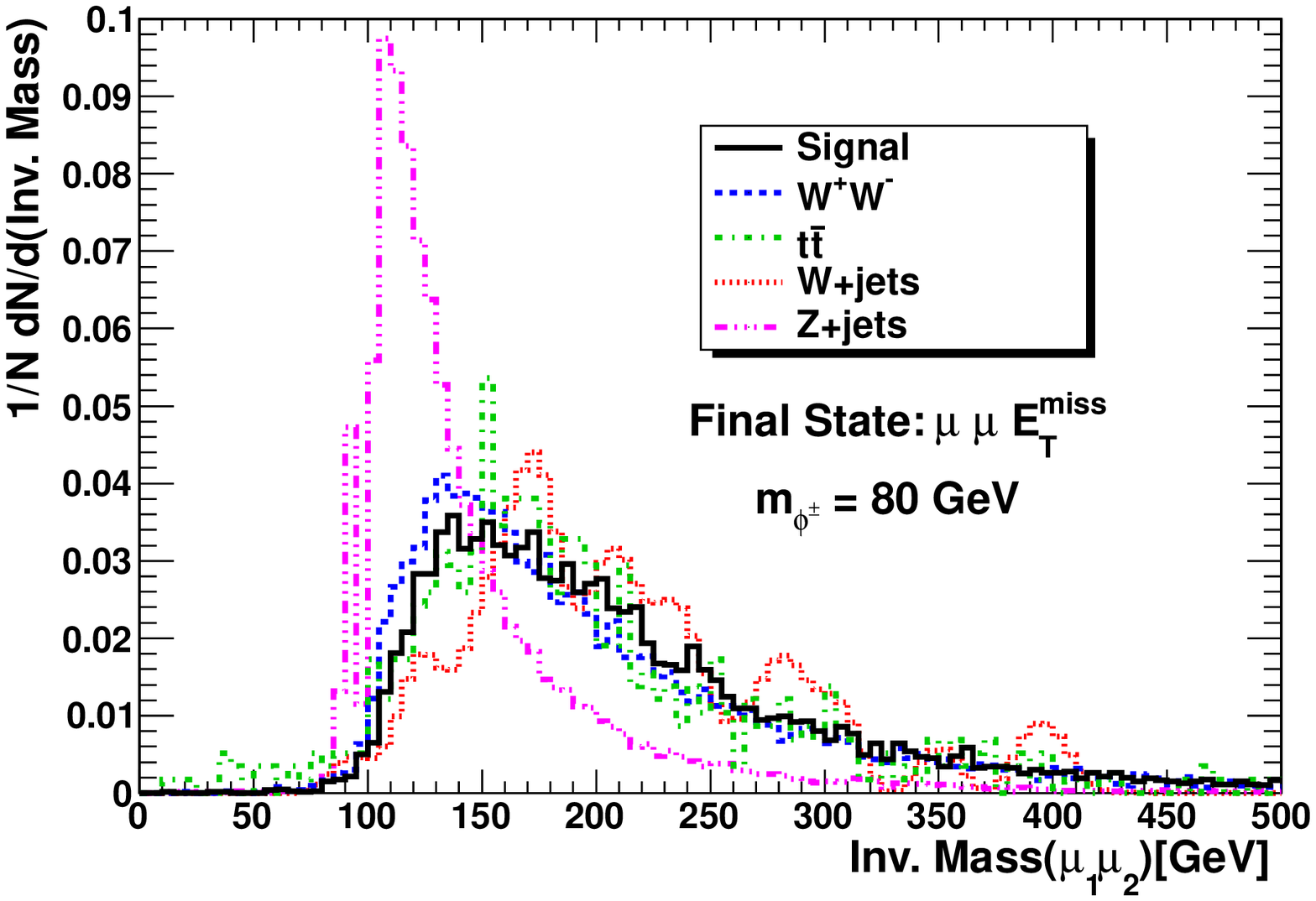}
\end{center}
\caption{Distribution of the invariant mass of the di-muon system in the $\mu\mu E^{miss}_{T}$ final state.\label{invmass}}
\end{figure}

\begin{figure}
\begin{center}
\includegraphics[width=0.80\textwidth]{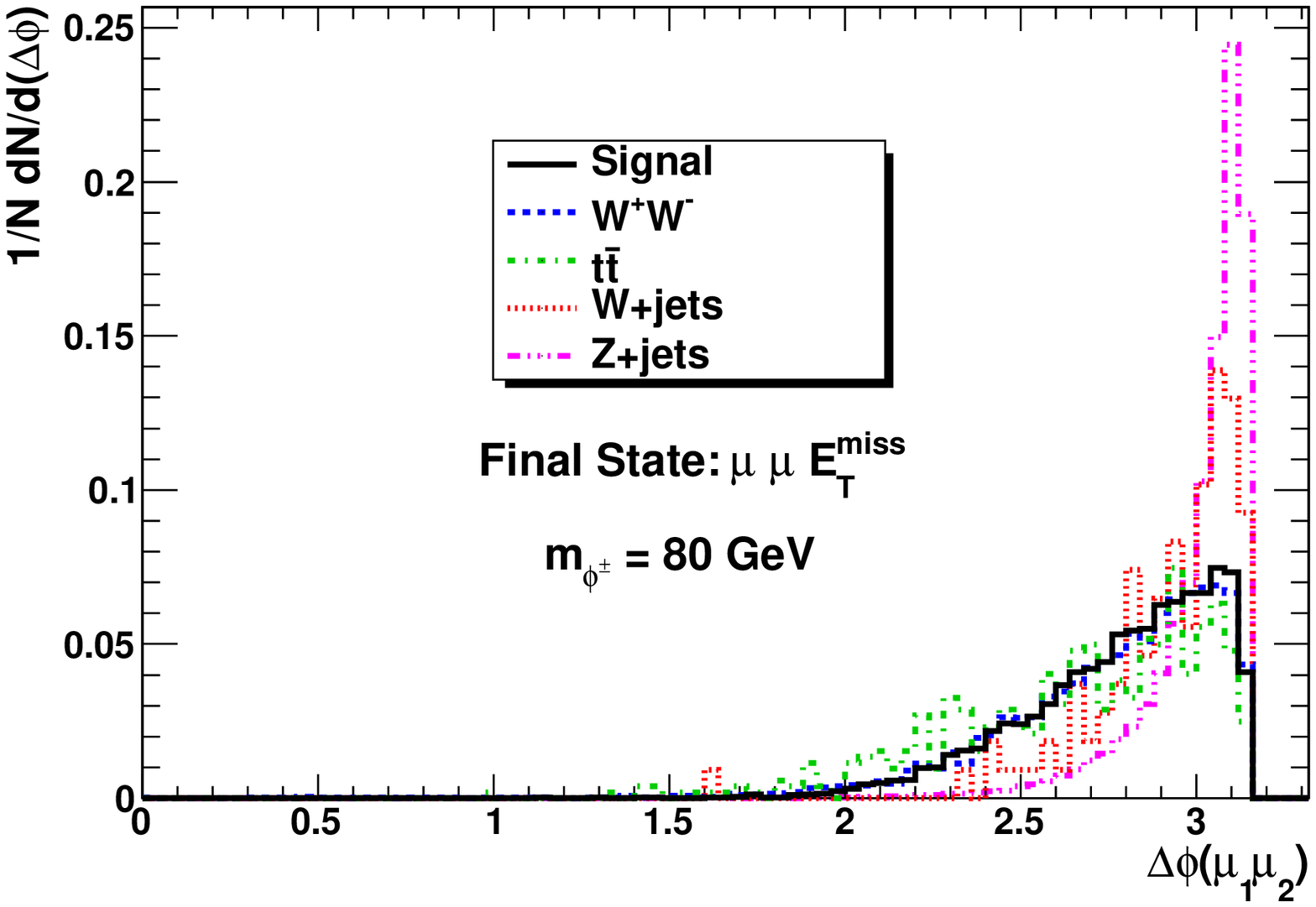}
\end{center}
\caption{The distribution of the azimuthal angle between the two
muons.\label{dphi2}}
\end{figure}

\begin{figure}
\begin{center}
\includegraphics[width=0.80\textwidth]{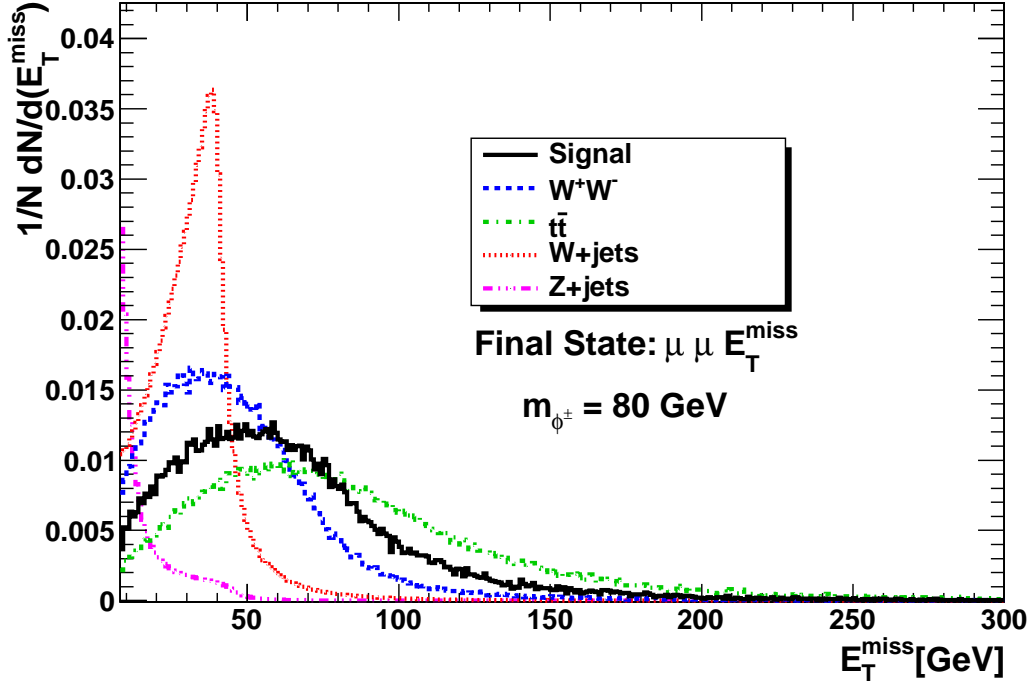}
\end{center}
\caption{Missing transverse energy distribution in the signal and
background events in the $\mu\mu E^{miss}_{T}$ final state.
\label{met2}}
\end{figure}

To search for this final state, the $\tau$ jet identification is
removed from the analysis. Instead, the focus is on muon
identification and $E^{miss}_{T}$ distributions. Figure
\ref{mupt2} shows the  transverse momentum distributions of the
muons. The muons in the events are required to satisfy the
following requirements:
\begin{equation}
p^{\mu}_{T}> 50~{\rm GeV},~ |\eta|<2.5 \ .
\label{mukin2}
\end{equation}
An event is required to have exactly two muons passing the above
requirements. As  seen from Fig. \ref{mupt2}, the soft muons in
the $Z$+jets events which are due to the off-shell production of
$Z^0$ bosons are removed by the above requirement and do not
contribute to the signal selection.
To study the number of jets in signal and background events, a jet
reconstruction is performed. Figure \ref{njets2} shows the number
of reconstructed jets which pass the following requirements:
\begin{equation}
 E^{\textnormal{jet}}_{T}>30 ~ \textnormal{GeV}, ~ |\eta|<2.5 \ .
\end{equation}
A jet is required to be separated enough from muons with $p_{T}>30$ GeV in the event with the following requirement:
\begin{equation}
\Delta R_{(\textnormal{jet,$\mu$})}>0.4 \ .
\end{equation}
Thus, to suppress the background events with jets in the final
state,
 it is required that no jet satisfying the above requirement
 exists in the event.
Since $Z$+jets events could be a potential background for $\mu\mu$
final state, the invariant mass of the two muons are plotted as
shown in Fig. \ref{invmass} and a cut on that is applied as the
following:
\begin{equation}
\textnormal{Inv. Mass} (\mu_{1},\mu_{2}) > 120~{\rm GeV} \ .
\end{equation}
In the next step the azimuthal angle between two muons is plotted
as shown in Fig. \ref{dphi2}.
In this analysis, we do not also apply any cut on $\Delta
\phi_{(\mu_1,\mu_2)}$. As  seen from Fig. \ref{dphi2}, the
$Z$+jets events tend to show stronger back-to-back muon pairs.
This effect can be due to the fact that muons in  the $Z$+jets
events come from a single mother particle. Further suppression of
this background is possible by applying a cut on this distribution
e.g. $\Delta \phi_{(\mu_1,\mu_2)}<3.$ However, since the final
signal significance exceeds 5$\sigma$, a study of the power of
this cut and further optimization of this search is left for a
full analysis including detector effects. Since the signal events
are produced with opposite charges of muons, the opposite charge
requirement is also applied to search for this final state:
\begin{equation}
\textnormal{charge}_{(\mu_1)}+\textnormal{charge}_{(\mu_2)} = 0\ .
\end{equation}
Finally the $E^{miss}_{T}$ distribution is obtained using Eq.
(\ref{meteq}).
\begin{equation}
E^{miss}_{T} =  |\sum_{i}{\overrightarrow{p_{Ti}}}| \ .
\label{meteq}
\end{equation}
The index $i$ runs over all stable particles in the event within the volume defined by $|\eta|<3.5$. This volume covers barrel and endcap calorimeters. The resulting distributions are shown in Fig. \ref{met2} and a cut on $E^{miss}_{T}$ is applied as the following:
\begin{equation}
E^{miss}_{T}>50~ \textnormal{GeV} \ .
\end{equation}
Table \ref{sigsel2} shows the selection efficiencies and the remaining number of signal events after applying each cut for different $m_{(\phi^{\pm})}$ hypotheses.

\begin{table}
\begin{center}
\begin{tabular}{|l|c|c|c|c|}
\hline
$m_{(\phi^{\pm})}$ & 80 GeV & 90 GeV & 110 GeV & 130 GeV  \\
\hline
Total cross section [fb] & 783 & 521 & 265 & 150  \\
\hline
Number of events at 30 $fb^{-1}$ & 5872 & 3907 & 1987 & 1125\\
\hline
N Muons = 2 & 1348(23$\%$) & 1080(28$\%$) & 720(36$\%$) & 494(44$\%$) \\
\hline
N Jets = 0 & 961(71$\%$) & 763(71$\%$) & 494(69$\%$) & 331(67$\%$) \\
\hline
Inv Mass($\mu,\mu$)$>$120 GeV & 895(93$\%$) & 711(93$\%$) &  463(94$\%$) & 308(93$\%$) \\
\hline
Opposite charge & 895(100$\%$) & 711(100$\%$) & 463(100$\%$) & 308(100$\%$) \\
\hline
$E^{miss}_{T}$ & 431(48$\%$) & 393(55$\%$) & 310(67$\%$) & 231(75$\%$) \\
\hline
total efficiency & 7.3$\%$ & 10.1$\%$ & 15.6$\%$ & 20.6$\%$ \\
\hline
Expected events at $30fb^{-1}$ & 431 & 393 & 310 & 231 \\
\hline
\end{tabular}
\end{center}
\caption{Selection efficiencies and remaining number of signal
events after each cut in the $\mu\mu E^{miss}_{T}$ final state.
Numbers in parenthesis are relative efficiencies in percent with
respect to the previous cut. Branching ratios have been taken
into account in transition from the second to third row.\label{sigsel2}}
\end{table}
As can be seen  the total selection efficiency
increases with increasing $m_{(\phi^{\pm})}$.
Table \ref{bsel2} shows selection efficiencies for background samples.
\begin{table}
\begin{center}
\begin{tabular}{|l|c|c|c|c|}
\hline
Process & $W^{+}W^{-}$ & $t\bar{t}$ & W+jets & Z+jets\\
\hline
Total cross section [pb] & 115.5 & 878.7 & 187.1$\times 10^3$ & 258.9$\times 10^3$\\
\hline
Number of events at 30 $fb^{-1}$ & 38713 & 2.9$\times10^5$  & 5.9$\times10^8$ & 2.6$\times10^8$\\
\hline
N Muons = 2 & 4120(11$\%$) & 52867(18$\%$) & 2635(4$\times10^{-4}\%$) & 752180(0.3$\%$) \\
\hline
N Jets = 0 & 2925(71$\%$) & 1705(3$\%$) & 1219(46$\%$) & 370599(49$\%$) \\
\hline
Inv Mass($\mu,\mu$)$>$120 GeV & 2620(90$\%$)& 1514(89$\%$) & 1160(95$\%$) & 220655(59$\%$) \\
\hline
Opposite charge & 2619(100$\%$) & 1447(96$\%$) & 1082(93$\%$) & 220544(100$\%$) \\
\hline
$E^{miss}_{T}$ & 730(28$\%$) & 937(65$\%$) & 295(27$\%$) & 234(0.1$\%$) \\
\hline
total efficiency & 1.9$\%$ & 0.3$\%$ & 5$\times10^{-5}\%$ & 9$\times10^{-5}\%$ \\
\hline
Expected events at $30fb^{-1}$ & 730 & 937 & 295 & 234\\
\hline
\end{tabular}
\end{center}
\caption{Selection efficiencies and remaining number of
background events after each cut in the $\mu\mu E^{miss}_{T}$ final
state. Numbers in parentheses are relative efficiencies in percent
with respect to the previous cut. Branching ratios have been taken
into account in transition from the second to third row.\label{bsel2}}
\end{table}
In this final state, $W$+jets background is more suppressed due to
requiring two muons in the event.
 While one of the muons can be genuine from the $ W$ decay,
  the other has to come from heavy meson decays appearing
  as a muon inside a jet cone. These muons are expected to
   be better suppressed with more sophisticated reconstruction and
   isolation algorithms used in the LHC full simulation analyses.
The $Z$+jets background is also expected to be under control with
the  cut on $E^{miss}_{T}$ and also the hard muon pair
requirement.
     As can be seen from Fig. \ref{mupt2}, the majority of these events
     tend to have either very soft muons or muons located around
     $E_{T}\simeq 45$ GeV (making the Z boson invariant mass peak).
      Thus, only a negligible fraction of these events appear with both
       muons harder than $E_{T}=50$ GeV. With the number of signal and
       background events as obtained in Tabs. \ref{sigsel2} and
       \ref{bsel2},
       the signal significance is calculated as shown in Table
       \ref{sig2}.
\begin{table}
\begin{center}
\begin{tabular}{|l|c|c|c|c|}
\hline
$m_{(\phi^{\pm})}$ & 80 GeV & 90 GeV & 110 GeV & 130 GeV \\
\hline
Signal significance & 9.2 & 8.4 & 6.6 & 4.9 \\
\hline
\end{tabular}
\end{center}
\caption{Signal significance in $\mu\mu E^{miss}_{T}$ final state for different $m_{(\phi^{\pm})}$ hypotheses.\label{sig2}}
\end{table}
As a rule of thumb, one can estimate the signal significance in
the $ee E^{miss}_{T}$ final state. For this search, the point B in
Table \ref{par} can be used. According to the values of
$g_{1\alpha}$
 which are equal for $\alpha=e,\mu$ and $\tau$, the branching ratios of
 the $\phi^{\pm}$ decays to different leptonic final states would be equal, {\it
  i.e.,}
  $BR(\phi^{\pm} \rightarrow e^\pm N_{1})=BR(\phi^{\pm} \rightarrow
  \mu^\pm N_{1})=BR(\phi^{\pm} \rightarrow \tau^\pm N_{1})\simeq 1/3$.
   Since the background samples have the same cross section times
branching ratios for both final states, one can use the signal
significance in $\mu\mu E^{miss}_{T}$ and scale the numerator of
Eq. \ref{signif} by a factor of 4/9 which leads to a significance
of 4.1$\sigma$ for $ee E^{miss}_{T}$ final state for
$m_{(\phi^{\pm})}=80$ GeV at 30 $fb^{-1}$. This estimation is
based on the assumption that the kinematics of events in $\mu\mu
E^{miss}_{T}$ and $ee E^{miss}_{T}$ final states are very similar.
From another point of view this is a simplistic estimation as
there is a difference between muon and electron reconstruction and
selection efficiencies in the detector. There can be also some
under-estimation due to the fact that in the $W$+jets events, the
electron fake rate is higher than the muon fake rate. A detailed
study of these features is beyond the scope of this paper and
needs a full simulation of the detector.
\section{Measuring the Model Parameters \label{parameters}}
In this section, an approach for measuring model parameters is
introduced and verified for the specific search described in this paper.\\
Having observed $N_{obs.}$ at 30$fb^{-1}$,
 the expected number of signal events has to be extracted from the
  observed number of events by subtracting the contamination of the background
   and then dividing by the total selection efficiency of the signal events
   which is obtained from the simulation. In other words,
\begin{equation}
 N_{S}=\frac{N_{obs.}-N_{B}}{\epsilon_S}
\end{equation}
where $\epsilon_S$ is the total selection efficiency of the
signal, $N_{S}$ is the expected number of signal events and
$N_{B}$ is the contamination in the observed sample which can be
derived either from real data or simulation. In the latter case,
the theoretical
 values of the backgrounds cross sections are used together with the
 selection efficiencies calculated from the simulation. Measuring
 $N_{S}$ in two different final states would allow possibility of
  comparing two numbers from the two final states and estimating branching ratio of $\phi^{\pm}$
 decays which in turn leads to the knowledge of model parameters
 such as $g_{i\alpha}$. \\
In order to estimate the uncertainty with which $N_{S}$ would be measured, the relative error on $N_{S}$ is calculated as the following:
 \begin{equation}
 \frac{\Delta N_S}{N_S}=\frac{\Delta\epsilon_S}{\epsilon_S}\oplus\frac{\Delta N_{obs.}}{N^{sel.}_S}\oplus\frac{\Delta N_B}{N^{sel.}_S}
\label{unc}
\end{equation}
where $\oplus$ means square root of quadratic sum of all terms.
Recall that $N_S$ is the original number of events produced at the
LHC,
 while $N^{sel.}_{S}$ is the number of selected signal events
 at the end of selection cuts; {\it i.e.,} $N^{sel.}_S=\epsilon_S N_S$.
  The relative uncertainty of the efficiency,
   ${\Delta\epsilon_S}/{\epsilon_S}$,
  is a quadratic sum of the main uncertainties; {\it i.e.,}
   the jet energy scale uncertainty, missing transverse energy
   uncertainty and lepton reconstruction uncertainty.
    As the best guesses for achievable uncertainties in LHC detectors,
    an uncertainty of 3$\%$ for jet energy scale Ref. \cite{jes} and
     missing transverse energy and 2$\%$ for lepton reconstruction and
      selection are assumed at an integrated luminosity of 30$fb^{-1}$.
      The total selection efficiency could therefore have an uncertainty of 5$\%$ or better which is smaller than the uncertainty of the background in the sample due to the large amount of remaining background events. Suppose $N_B$ is measured using theoretical cross
 section values and selection efficiencies from simulation. The
 uncertainty of $N_B$ relative to signal is then
\begin{equation}
\frac{\Delta N_B}{N^{sel.}_S}=(\frac{\Delta\sigma}{\sigma}\oplus
\frac{\Delta L}{L}\oplus\frac{\Delta\epsilon_B}{\epsilon_B})
\frac{N_B}{N^{sel.}_S}\ .
\end{equation}

The uncertainty of the theoretical background cross sections
originates from the choice of PDF set, the PDF set parameters
errors, the value of $\alpha_{S}$ used in the PDF set and the
renormalization and factorization scale uncertainty. Some of these
uncertainties have been estimated for the $t\bar{t}$ events in
Ref. \cite{ttbarXsec} and lead to a $\sim 14\%$ uncertainty for
this background which should be added to the uncertainty due to
the PDF fit parameters. We also study the theoretical background
uncertainties independently using MCFM package (excluding the
error from PDF fit parameters which needs a dedicated detailed
analysis). Results are shown in Tab. \ref{bunc}.
\begin{table}
\begin{center}
\begin{tabular}{|l|c|c|c|c|}
\hline
Channel & $WW$ & W+jets & Z+jets & $t\bar{t}$ \\
\hline
Central Cross Section & 115.5 pb & 187.1 nb & 258.9 nb & 878.6 pb\\
\hline
Ren. Scale $\times$ 2 & 111.5(-3.5$\%$) & 191.7(2.1$\%$) & 263.1(1.6$\%$) & 777.2(-11.5$\%$)\\
\hline
Ren. Scale $\div$ 2 & 117.7(1.9$\%$) & 181.8(-2.8$\%$) & 252.2(-2.6$\%$) & 977.3(11.2$\%$)\\
\hline
NNPDF2.0 & 114.5(-0.9$\%$) & 185.9(-0.6$\%$) & 285.5(10.3$\%$) & 917.6(4.4$\%$)\\
\hline
MSTW2008nnlo & 119.5(3.5$\%$) & 198.1(5.9$\%$) & 270.9(4.6$\%$) & 861.9(-1.9$\%$)\\
\hline
CTEQ6.6 & 113.3(-1.9$\%$) & 196.3(4.9$\%$) & 302.9(17$\%$) & 841.7(-4.2$\%$)\\
\hline
$\alpha_{S}$=0.1127 & 114.3(-1$\%$) & 186.2(-0.5$\%$) & 262.2(1.3$\%$)& 813.6(-7.4$\%$)\\
\hline
$\alpha_{S}$=0.1207 & 116.9(1.2$\%$) & 187.9(0.4$\%$) & 255.7(-1.2$\%$) & 940.7(7$\%$)\\
\hline
Total uncertainty & 4.6$\%$ & 6.4$\%$ & 17$\%$ & 14$\%$\\
\hline
\end{tabular}
\end{center}
\caption{Theoretical uncertainty of background processes excluding the uncertainty arising from the PDF set parameters errors. The central value of cross sections has been calculated using MRST2004 NNLO PDF set with a central value of $\alpha_{S}=0.1167$. \label{bunc}}
\end{table}
The renormalisation scale is set to the characteristic mass in the
event, {\it i.e.,} the top quark, $W$ boson and $Z$ boson masses
for $t\bar{t}$, $W$+jets, $WW$ and $Z$+jets respectively. The
variation of renormalisation scale is then performed around the
central value. The central value of $\alpha_{S}$ according to PDG
Ref. \cite{pdgalphas} is 0.1176$\pm$0.002. The central value of
this parameter chosen by different PDF sets is different, however,
the 0.002 uncertainty is interpreted as one-$\sigma$ or 68$\%$
uncertainty and  a 95$\%$ uncertainty ($\sim2\sigma$) equal to
0.004 is therefore used for the $\alpha_{S}$ variation. This leads
to the values of 0.1127 and 0.1207 using the central $\alpha_{S}$
value adopted by MRST2004 NNLO PDF set. The total error has been
calculated as proposed in Eq. \ref{toterr}.
\begin{equation}
\textnormal{Total error}=
\sqrt{\frac{(\textnormal{Ren. Scale}\uparrow)^{2}+
(\textnormal{Ren. Scale}\downarrow)^{2}}{2}+
(\textnormal{max~PDF~set~error})^{2}+\frac{(\alpha_{S}\uparrow)^{2}+
(\alpha_{S}\downarrow)^{2}}{2}}
\label{toterr}
\end{equation}
where $(\textnormal{Ren. Scale}\uparrow)$ and $(\textnormal{Ren.
Scale}\downarrow)$ are the observed error on the central cross
sections when the renormalisation scale is  respectively set to
twice and half the central value. Similarly,
$(\alpha_{S}\uparrow)$ and $(\alpha_{S}\downarrow)$ are errors on
central cross sections when $\alpha_S$ is set to respectively
0.1207 and 0.1127. For example, for the $WW$ case,
$(\textnormal{Ren. Scale}\uparrow)=-3.5~ \%$, $(\textnormal{Ren.
Scale}\downarrow)=1.9~\%$, $(\alpha_{S}\uparrow)=1.2~\%$ and
$(\alpha_{S}\downarrow)=-1~ \%$.
 The middle term in Eq. (\ref{toterr}) is the maximum uncertainty
  observed due to a
specific PDF set.
As the error due to the PDF fit parameters can be of the order of the
error due to the choice of PDF or $\alpha_{S}$ variation
 (e.g. cf. Ref. \cite{pdfalphas}) an average total error of $\sim 10 \%$
is expected for the main background processes. It should be noted
that here the errors due to the top quark, $W$ and $Z$ boson mass
measurements were neglected although they could also contribute
sizably
to the total error. 
All arguments above imply that an average 10$\%$ error on the main
 backgrounds is a reasonable assumption.\\

Taking 3$\%$ uncertainty for the LHC luminosity  \cite{lumi}
 and 10$\%$ uncertainty on background cross sections,
 for $m_{(\phi^{\pm})}=80$ GeV, this term is estimated to be
\begin{equation}
\frac{\Delta N_B}{N^{sel.}_S}(\mu\mu)\simeq 58\% \ .
\end{equation}
Here only the result on the $\mu\mu$ final state  has been
presented as the $\tau\mu$ final state suffers from the large
background contamination.

The second term on the right side of Eq. (\ref{unc}) refers to the
ratio of statistical uncertainty of the observed number of events
to the number of signal events and can be written as
\begin{equation}
\frac{\Delta N_{obs.}}{N^{sel.}_S}=\frac{\sqrt{N_{obs.}}}{N^{sel.}_S}.
\end{equation}
This term amounts to 12$\%$ for the $\mu\mu$ final states. Thus,
the dominant contribution to $\Delta N_S/N_S$ is the last term of
Eq. (\ref{unc}).
As far as the uncertainties in the selection of events and the
collider luminosity and cross section values remain the same, the
last term, which is proportional to ${N_B}/{N^{sel.}_S}$, does not
change with luminosity. An improvement is therefore  possible only
by acquiring a better knowledge of the uncertainties involved in
the analysis or increasing the center of mass energy to obtain a
higher ratio of signal to background. In principle, by using the
data from the LHC itself, it would be possible to reduce the
uncertainties.
\section{\label{alternative}An Alternative Channel: the $\phi^{\pm}\phi_{2}+\phi^{\pm}\delta_{2}$ Production}
\begin{figure}
\begin{center}
\includegraphics[width=0.80\textwidth]{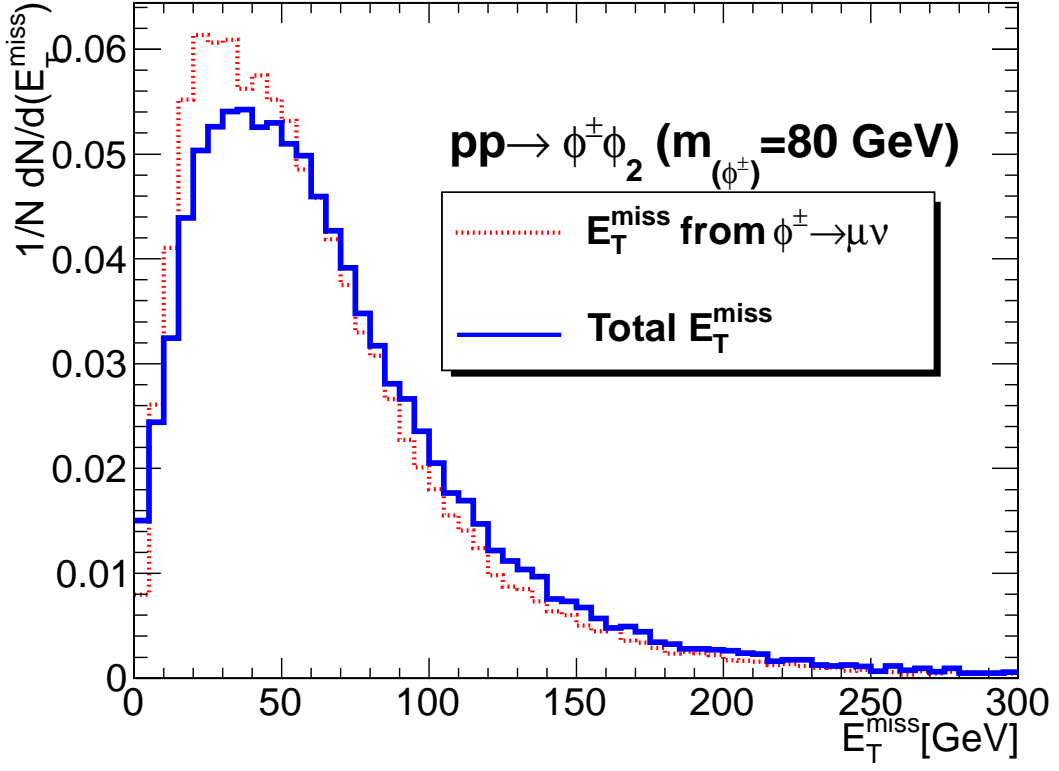}
\end{center}
\caption{Distribution of total $E^{miss}_{T}$ and its comparison
with  distribution of $E^{miss}_{T}$ from $\phi^{\pm}\rightarrow
\mu \nu$.\label{metcompare}}
\end{figure}

\begin{figure}
\begin{center}
\includegraphics[width=0.80\textwidth]{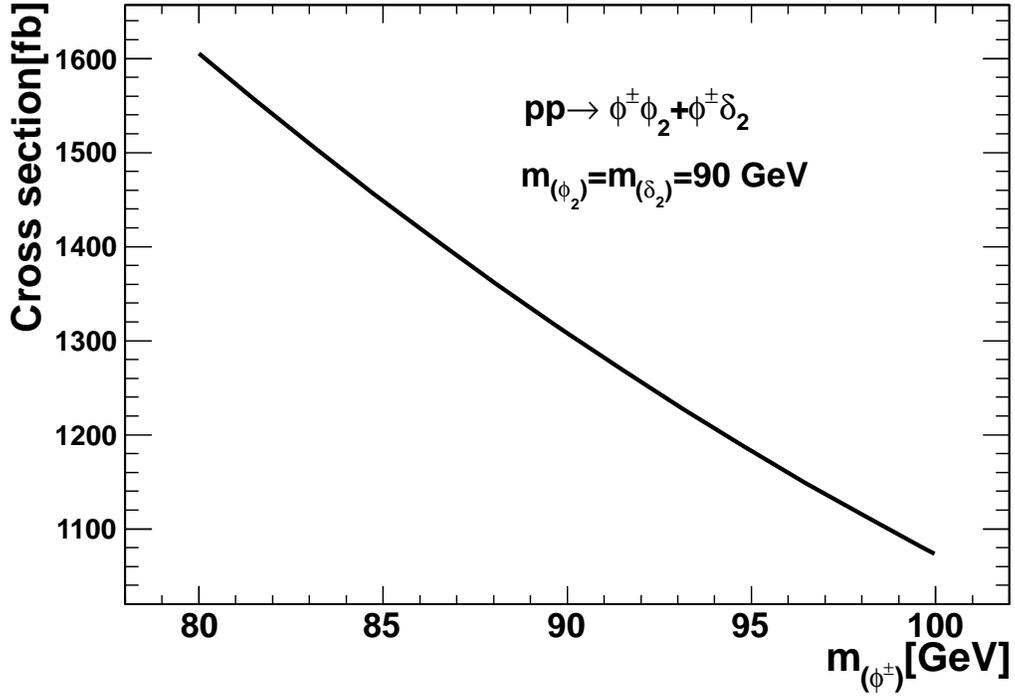}
\end{center}
\caption{Total signal cross section in the $\phi^{\pm}\phi_{2}$ channel as a function of $\phi^{\pm}$ mass.\label{xsecAH}}
\end{figure}

\begin{figure}
\begin{center}
\includegraphics[width=0.80\textwidth]{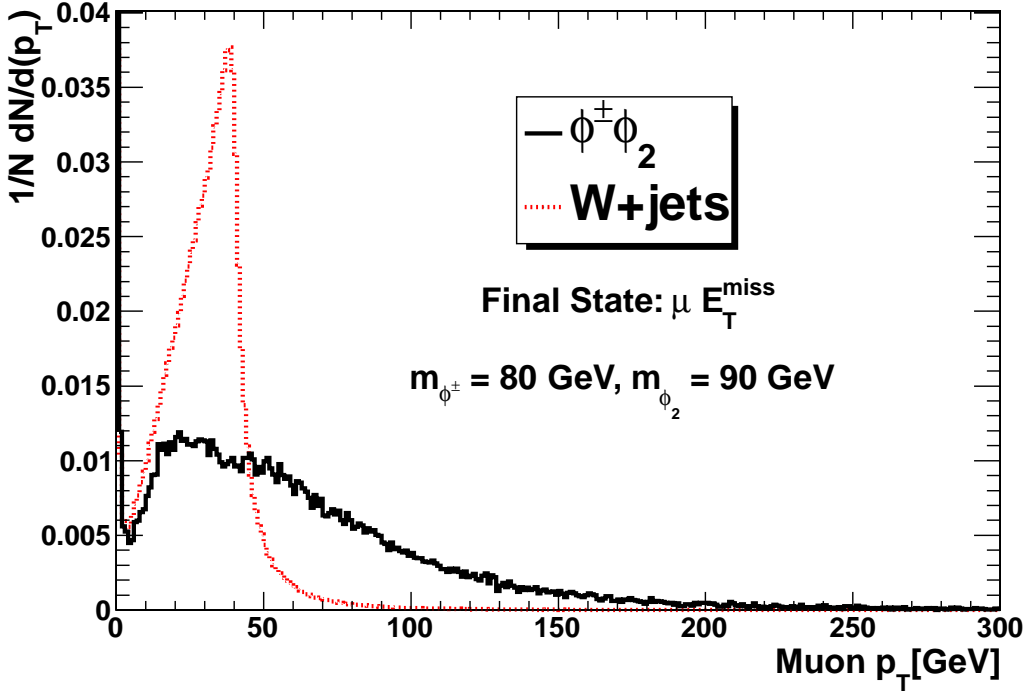}
\end{center}
\caption{Muon transverse momentum distribution in $\phi^{\pm}\phi_{2}$ events in $\mu E^{miss}_{T}$ final state.\label{muonpt3}}
\end{figure}

\begin{figure}
\begin{center}
\includegraphics[width=0.80\textwidth]{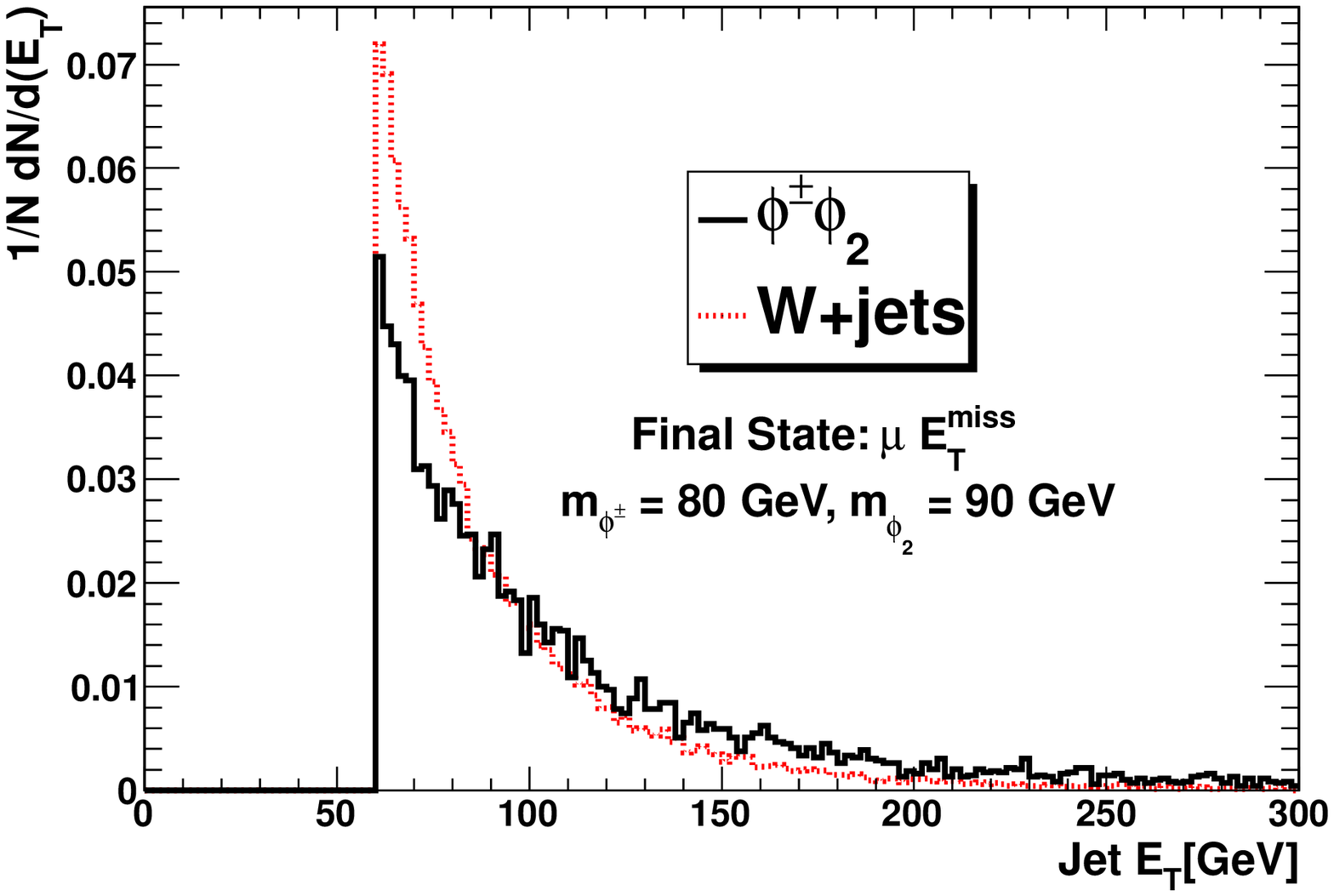}
\end{center}
\caption{Jet transverse energy distribution in
$\phi^{\pm}\phi_{2}$ events in $\mu E^{miss}_{T}$ final state.
Jets satisfying Eq. \ref{jeteteq2} fill the histogram. Softer jets
with $E_T<60$ GeV are skipped due to the same reason  explained in
caption of Fig. \ref{jetet}.\label{jetet3}}
\end{figure}

\begin{figure}
\begin{center}
\includegraphics[width=0.80\textwidth]{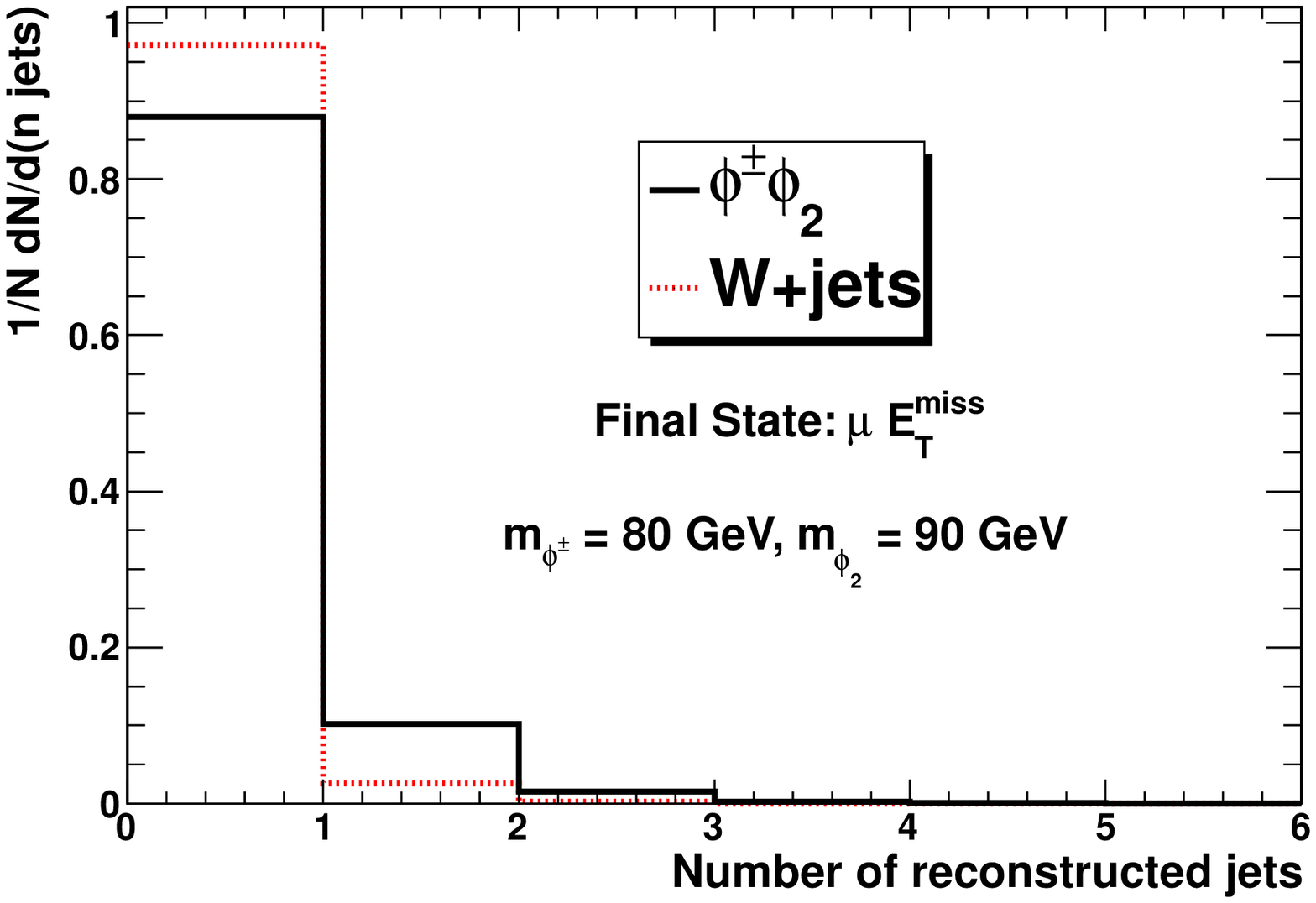}
\end{center}
\caption{Number of reconstructed jets for the $\phi^{\pm}\phi_{2}$
events in $\mu E^{miss}_{T}$ final state. A jet is considered and
counted if it passes the kinematic requirements described in the
text.\label{jetmul3}}
\end{figure}

\begin{figure}
\begin{center}
\includegraphics[width=0.80\textwidth]{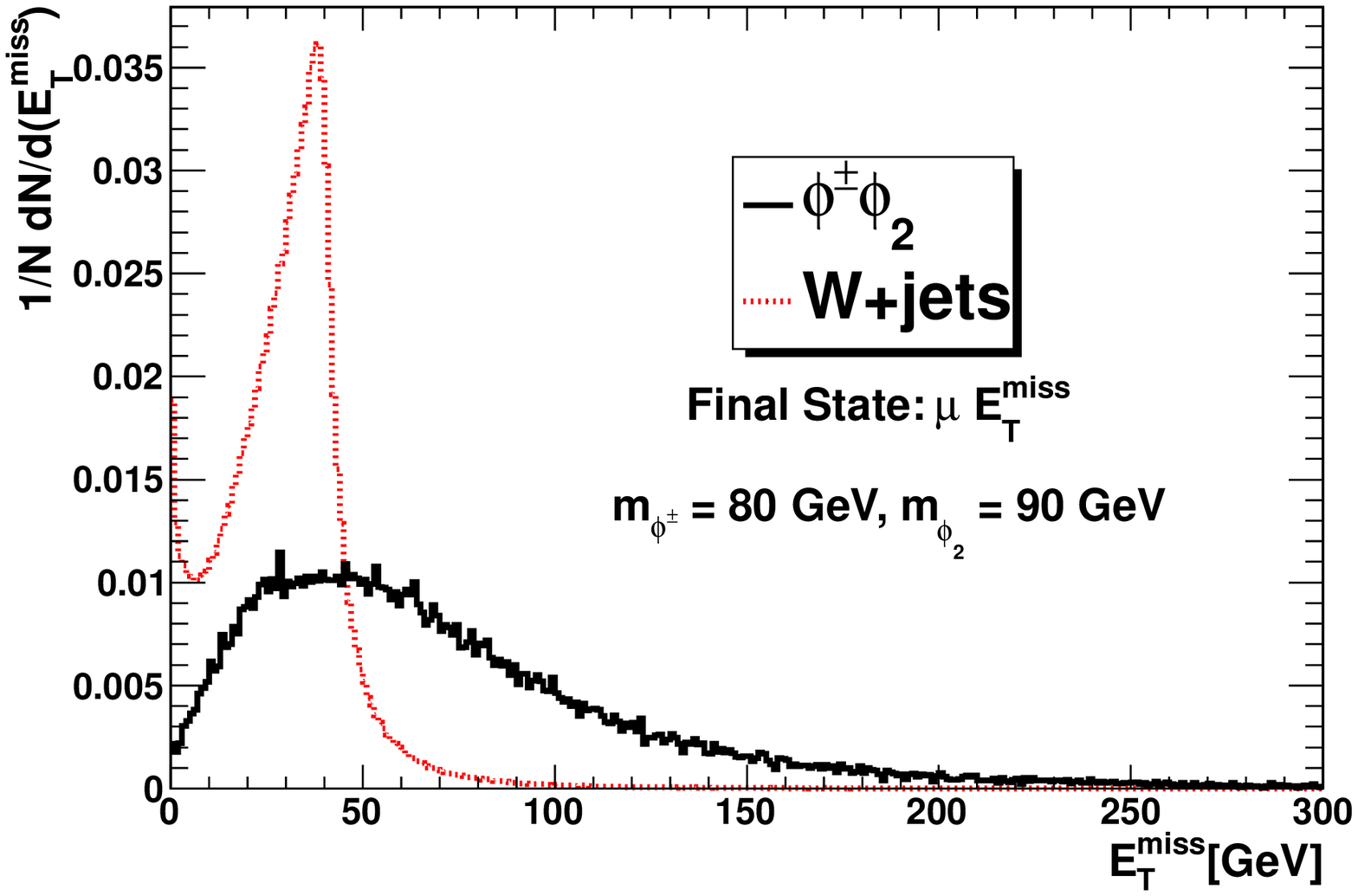}
\end{center}
\caption{Missing transverse energy distribution for the
$\phi^{\pm}\phi_{2}$ events in the $\mu E^{miss}_{T}$ final
state.\label{met3}}
\end{figure}

\begin{figure}
\begin{center}
\includegraphics[width=0.80\textwidth]{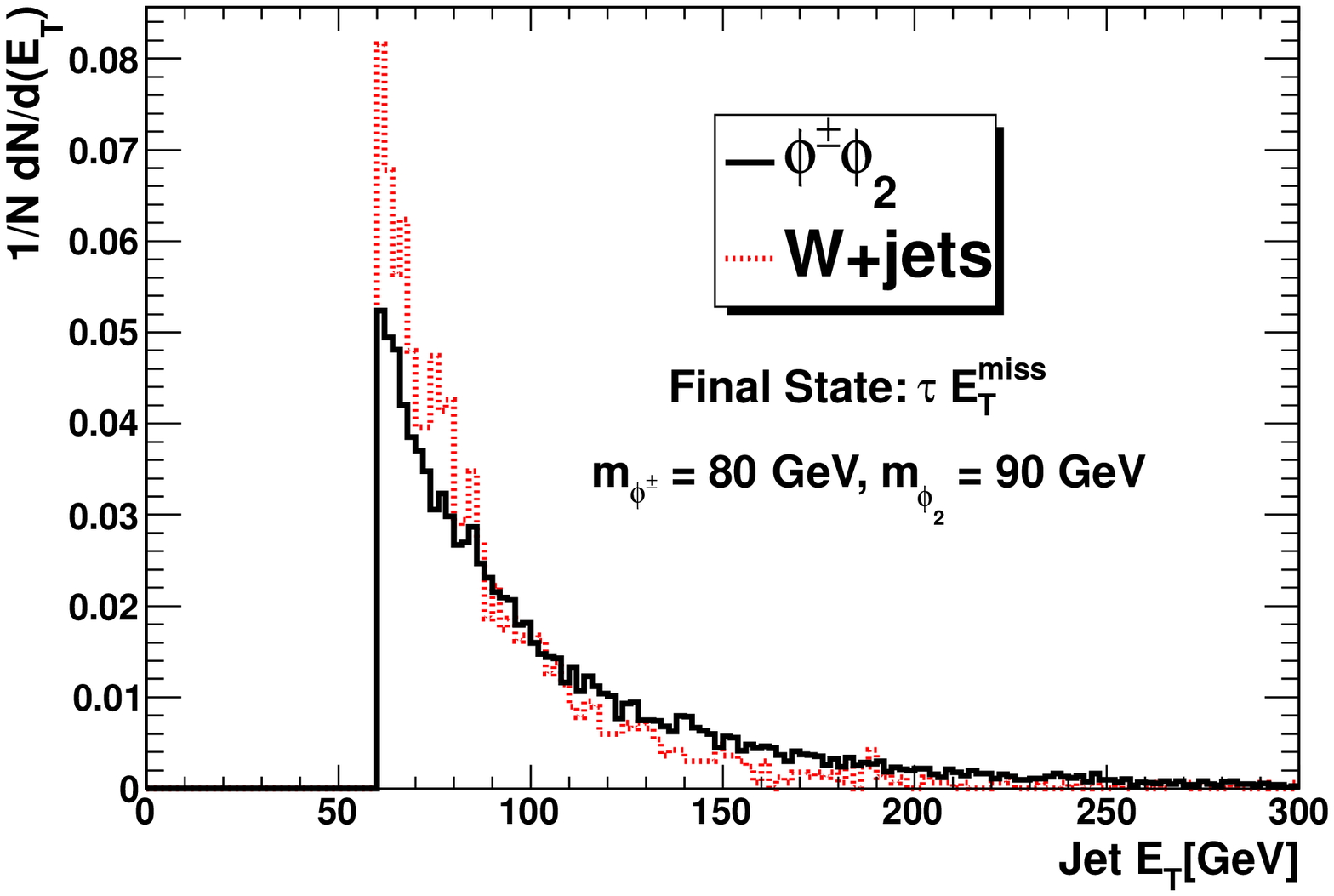}
\end{center}
\caption{Jet transverse energy distribution for the
$\phi^{\pm}\phi_{2}$ events in the $\tau E^{miss}_{T}$ final
state. Here also only jets satisfying Eq. \ref{jeteteq2} fill the
histogram. Softer jets with $E_T<60$~GeV are skipped due to the
same reason explained in caption of Fig. \ref{jetet}.
\label{jetet4}}
\end{figure}

\begin{figure}
\begin{center}
\includegraphics[width=0.80\textwidth]{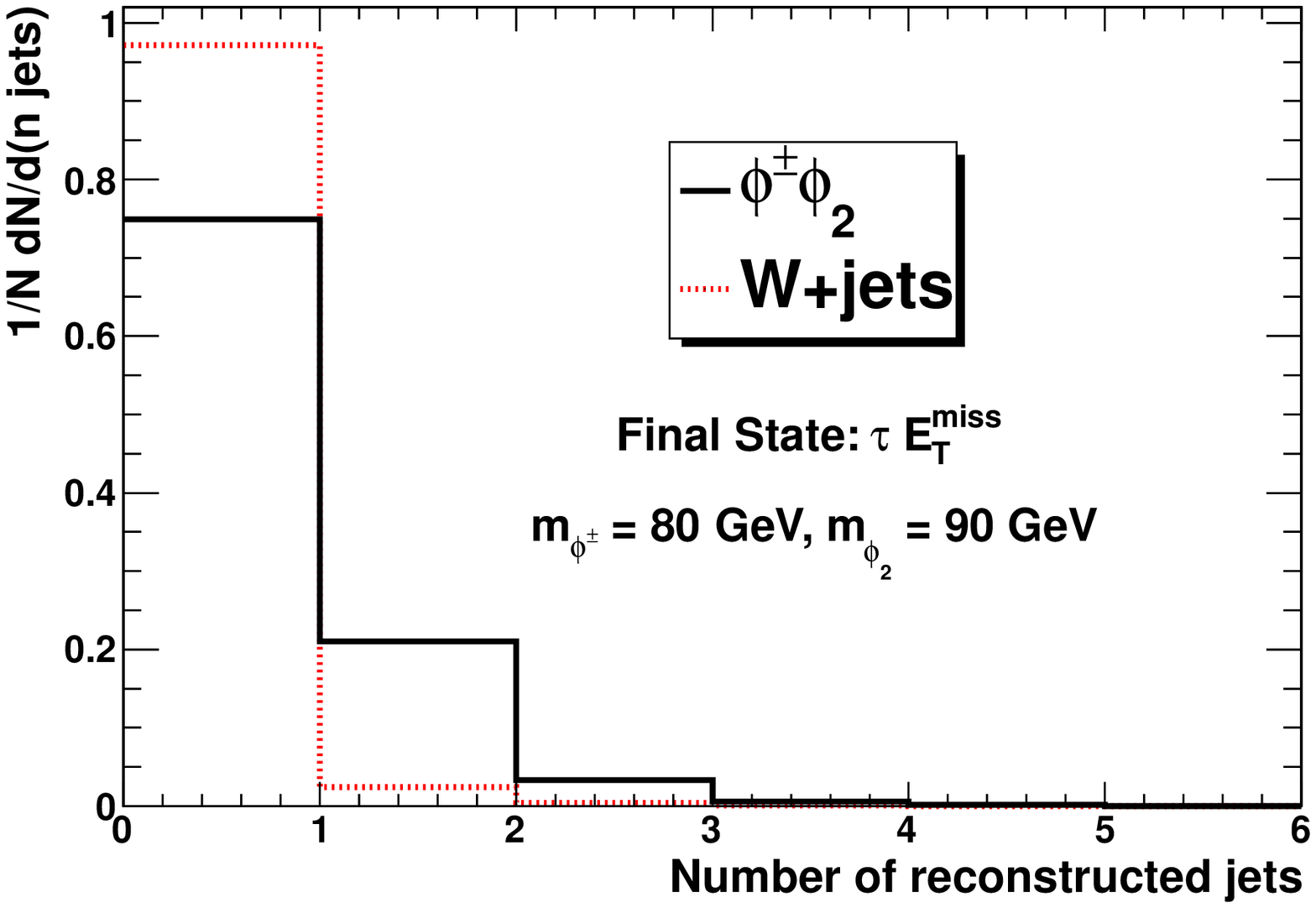}
\end{center}
\caption{Number of reconstructed jets for the $\phi^{\pm}\phi_{2}$
events in the $\tau E^{miss}_{T}$ final state. A jet is considered
and counted if it passes the kinematic requirements described in
the text.\label{jetmul4}}
\end{figure}

In this section, we study  the alternative channel, $pp\rightarrow
\phi^{\pm}\phi_{2}+\phi^{\pm}\delta_{2}$. Remember that the gauge
coupling of the neutral CP-odd and CP-even components of $\Phi$
($\phi_1$ and $\phi_2$) to $W^\pm \phi^\mp$ are similar. On the
other hand, for small values of the mixing ($\alpha \simeq 0.01$),
 $\delta_2\simeq\phi_2$.
 From the production
point of view, $\delta_2$ therefore behaves like $\phi_2$ so
 it is enough to study the first process ({\it i.e.,} $pp\rightarrow \phi^{\pm}\phi_{2}$)
  and double the number of events to account for the
$\phi^{\pm}\delta_2$
  production as well.
In this analysis, we take $m_{\phi_2}=m_{\delta_2}=90~$GeV in
accord with Ref. \cite{neutralHiggs} which imposes a lower limit
of 90 GeV on the neutral Higgs boson mass.
 With the above assumptions, processes such as
$\phi_{2},\delta_{2}\rightarrow\phi^{\pm}W^{\mp}$,
 $\phi_{2}\rightarrow\delta_{2}Z^{0}$ or
 $\delta_{2}\rightarrow\phi_{2}Z^{0}$ will be forbidden and
  $\phi_{2}(\delta_{2})$ will have only invisible decay modes.
  Since $\phi_{2}$ appears as missing energy in the
  detector, the final state to analyze is $\mu E^{miss}_{T}$ or
  $\tau E^{miss}_{T}$ depending on the decay of $\phi^{\pm}$ to
  a muon or $\tau$. As before, we take
$\textnormal{BR}(\phi^{\pm}\rightarrow \mu^\pm N_1)=
\textnormal{BR}(\phi^{\pm}\rightarrow \tau^\pm N_1) = 0.5$.\\
The simulation of these events is done by considering
 that $\phi_{2}$,
 a CP-odd neutral scalar, would behave like $A^{0}$ 
 in MSSM and its production process would therefore resemble
  the $H^{\pm}A^{0}$ production in the framework of the MSSM.
Within the MSSM, the CP-odd and CP-even Higgs particles are
produced similarly.  Since the  $H^{\pm}A^{0}$ process
 is not produced by PYTHIA, we use $H^{\pm}H^{0}$ to produce the signal
 events. The coupling of $H^{\pm}$ to $W^{\pm}H^{0}$ is given
by $({e^{2}}/{\sin^{2}\theta_{W}})\sin(\beta-\alpha)$ so setting
  $\cos(\beta-\alpha)=0$,  the coupling will be the same
as the coupling of $\phi^\pm$ to $W^\mp \delta_2$ within our model.
   Moreover, in this
   analysis  we  set $\tan(\beta)=100$ and obtain cross sections in
   agreement with Ref. \cite{A0H+}.
 To produce the same type of events, in production of $H^{\pm}H^{0}$
 events, $H^{0}$ is forced to decay only to a muon pair and those muons
 are ignored in the event to have the same lepton multiplicity as in
  $\phi^{\pm}\phi_{2}$ production. The missing transverse energy is
  also
  calculated by ignoring the $H^{0}$ products (muons) in the event
  so
   treating it as a source of invisible energy. Figure \ref{metcompare}
    compares $E^{miss}_{T}$ as calculated only from the $N_1$ in
     $\phi^{\pm}$ decay and the total
  $E^{miss}_{T}$ including invisible energy from $\phi_{2}$.
Figure \ref{xsecAH} shows the calculated total cross section
 of the signal for different
$\phi^{\pm}$ masses. This is the inclusive cross section although
there is a little difference between the production rate of
$\phi^{+}$ and $\phi^{-}$ at the LHC.
Since the final state is either $\mu E^{miss}_{T}$ or $\tau E^{miss}_{T}$,
the main background for this signal would be single $W$ boson production.
 All other backgrounds are expected to be suppressed by requirements on
 lepton multiplicity and central jet veto.
Figures \ref{muonpt3}, \ref{jetet3}, \ref{jetmul3} and \ref{met3} compare
the signal and background distributions of the muon transverse momentum,
jet transverse energy, jet multiplicity and $E^{miss}_{T}$ respectively
in the $\mu E^{miss}_{T}$ final state.
Based on these distributions and since the background cross section is high and needs a large suppression factor, the following kinematic cuts are applied:
\begin{equation}
 \textnormal{Muon} ~ p_{T} > 100 ~\textnormal{GeV},~ |\eta|<2.5\ ,
\end{equation}
\begin{equation}
 E^{miss}_{T} > 100 ~\textnormal{GeV} \ .
\end{equation}
\begin{equation}
 \textnormal{Jet}~ E_{T} > 60 ~\textnormal{GeV}, ~|\eta|<2.5 \ .
\label{jeteteq2}
\end{equation}
There should be no jet in the event with $\mu E^{miss}_{T}$ final state passing the above requirement. Similarly Figs. \ref{jetet4} and \ref{jetmul4} compare the signal
and background distributions of the jet transverse energy and jet
multiplicity respectively in the $\tau E^{miss}_{T}$ final state.
The missing transverse energy distribution is not studied based on
the same argument as mentioned in the $\tau \mu E^{miss}_{T}$
final state analysis of $\phi^{+}\phi^{-}$ process. For this final
state the same cut on the jet $E_{T}$ is applied. Since there is a
$\tau$ lepton in the event, the same $\tau$-jet identification as
in the analysis of $\tau \mu E^{miss}_{T}$ is applied. Similarly
to the $\tau \mu E^{miss}_{T}$ case, the polarization dependent
cuts are not applied.
Based on selection requirements of the two final states, Tabs. \ref{selmu}
and \ref{seltau} show selection efficiencies of signal and background
events and the remaining number of events.
\begin{table}
\begin{center}
\begin{tabular}{|l|c|c|}
\hline
Final state & \multicolumn{2}{|c|}{$\mu E^{miss}_{T}$} \\
\hline
Process & Signal  & W+jets \\
\hline
Total cross section [pb] & 1604 & 187.3$\times 10^{3}$ \\
\hline
Number of events at 30 $fb^{-1}$ & 24057 & 5.9$\times 10^{8}$ \\
\hline
N Muons = 1 & 3928(16$\%$) & 1952900(0.3$\%$) \\
\hline
N Jets = 0 & 3182(81$\%$) & 356404(18$\%$)\\
\hline
$E^{miss}_{T}>100$ GeV & 2652(83$\%$) & 87319(24$\%$) \\
\hline
total efficiency & 11$\%$ & 0.015$\%$ \\
\hline
Expected events at $30fb^{-1}$ & 2652 & 87319 \\
\hline
\end{tabular}
\end{center}
\caption{Selection efficiencies and remaining number of background
events after each cut in the $\mu E^{miss}_{T}$ final state.
Numbers in parentheses are relative efficiencies in percent with
respect to the previous cut. Branching ratios have been taken into
account in transition from the third to forth row. To calculate
the signal, both $\phi^{\pm}\phi_{2}$ and $\phi^{\pm}\delta_{2}$
have been taken into account. \label{selmu}}
\end{table}
\begin{table}
\begin{center}
\begin{tabular}{|l|c|c|}
\hline
Final state & \multicolumn{2}{|c|}{$\tau E^{miss}_{T}$} \\
\hline
Process & Signal  & W+jets \\
\hline
Total cross section [pb] & 1604 & 187.3$\times 10^{3}$ \\
\hline
Number of events at 30 $fb^{-1}$ & 24057 & 5.9$\times 10^{8}$ \\
\hline
N Jets = 1 & 5117(21.3$\%$) & 1.5e+07(2.6$\%$) \\
\hline
Isolation & 3058(59.8$\%$) & 1.2e+06(7.7$\%$)\\
\hline
1- or 3-prong decay & 2900(94.8$\%$) & 389380(33$\%$) \\
\hline
total efficiency & 12.05$\%$ & 0.066$\%$ \\
\hline
Expected events at $30fb^{-1}$ & 2900 & 389380 \\
\hline
\end{tabular}
\end{center}
\caption{Selection efficiencies and remaining number of background
events after each cut in the $\tau E^{miss}_{T}$ final state.
Numbers in parentheses are relative efficiencies in percent with
respect to the previous cut. Branching ratios have been taken into
account in transition from the third to forth row. To calculate
the signal,  both $\phi^{\pm}\phi_{2}$ and $\phi^{\pm}\delta_{2}$
have been taken into account.\label{seltau}}
\end{table}
Thus for $m_{(\phi^{\pm})}=80$ GeV, a signal significance of
9$\sigma$ and 4.6$\sigma$ is respectively obtained for $\mu
E^{miss}_{T}$ and $\tau E^{miss}_{T}$ final states at 30 $fb^{-1}$
integrated luminosity. Choosing the other point in the parameter
space (the point B in Table \ref{par}), one obtains $BR(\phi^{\pm}
\rightarrow e^\pm N_{1})=1/3$ so the signal cross section in the
$e E^{miss}_{T}$ final state would be two third of that in the
$\mu E^{miss}_{T}$ final state using point A in Table \ref{par}.
The $e E^{miss}_{T}$ significance is therefore expected to be $2/3
\times 9 = 6\sigma$. Since the
 cross section of the signal decreases with increasing
 $m_{(\phi^{\pm})}$,
higher $\phi^{\pm}$ masses would lead to lower signal significances.

\section{Discovery Potential at the 7 TeV Run\label{discovery}}
In this section, an estimation of signal significance at different
final states is made by a simple rescaling of the signal and
background
 cross sections to their corresponding values at the 7 TeV center
  of mass
 energy. Although a detailed study of kinematic distributions at this
energy is needed to reach a concrete conclusion, a simple cross
section rescaling would give  first feeling on how signal
significance would
 decrease at lower center of mass energies. We choose the
 7 TeV center of
  mass
  energy to discuss  the possibility of establishing
  the model introduced
in this paper at the current LHC run. Table \ref{ratios} shows ratio
of
7 TeV to 14 TeV cross section of signal and background processes.
\begin{table}
\begin{center}
\begin{tabular}{|l|c|c|}
\hline
\multicolumn{3}{|c|}{Signal} \\
\hline
Channel & Mass Point & 7 TeV to 14 TeV Ratio \\
\hline
\multirow{4}{*}{$\phi^{+}\phi^{-}$} &
$m_{(\phi^{\pm})}=80$ GeV & 0.4 \\
& $m_{(\phi^{\pm})}=90$ GeV & 0.39 \\
& $m_{(\phi^{\pm})}=110$ GeV & 0.37 \\
& $m_{(\phi^{\pm})}=130$ GeV & 0.35 \\
\hline
$\phi^{\pm}\phi_{2}$ & $m_{(\phi^{\pm})}=80$ GeV & 0.4 \\
\hline
\hline
\multicolumn{3}{|c|}{Background} \\
\hline
\multicolumn{2}{|c|}{Channel} & 7 TeV to 14 TeV Ratio \\
\hline
\multicolumn{2}{|c|}{$W^{+}W^{-}$} & 0.38 \\
\hline
\multicolumn{2}{|c|}{$t\bar{t}$} & 0.19 \\
\hline
\multicolumn{2}{|c|}{W+jets} & 0.49 \\
\hline
\multicolumn{2}{|c|}{Z+jets} & 0.6 \\
\hline
\end{tabular}
\end{center}
\caption{Ratio of 7 TeV to 14 TeV cross sections of signal and background events.\label{ratios}}
\end{table}
These ratios lead to the  signal significance
 as shown in Table \ref{7TeVsig} for different final states, provided
 that the data corresponding to 30 $fb^{-1}$ integrated luminosity is
 collected at the 7 TeV center of mass energy. This assumption might be in
 contrast to the current plan of the LHC machine, as there may be a
 switch to higher center of mass energies (10 TeV and 14 TeV later on)
 before this amount of data is collected. If this phase finishes
 after collecting only 1 $fb^{-1}$, the expected signal
 significance will
 be $1/\sqrt{30}\simeq 0.18$ times smaller so there will be no
 discovery chance.
\begin{table}
\begin{center}
\begin{tabular}{|l|c|c|}
\hline
Channel & Mass Point & Signal significance \\
\hline
\multirow{4}{*}{$\phi^{+}\phi^{-}\rightarrow \tau\mu E^{miss}_{T}$ } &
$m_{(\phi^{\pm})}=80$ GeV & 1.6\\
& $m_{(\phi^{\pm})}=90$ GeV & 1.2 \\
& $m_{(\phi^{\pm})}=110$ GeV & 0.7 \\
& $m_{(\phi^{\pm})}=130$ GeV & 0.5 \\
\hline
\multirow{4}{*}{$\phi^{+}\phi^{-}\rightarrow \mu\mu E^{miss}_{T}$ } &
$m_{(\phi^{\pm})}=80
$ GeV & 6.4 \\
& $m_{(\phi^{\pm})}=90$ GeV & 5.7 \\
& $m_{(\phi^{\pm})}=110$ GeV & 4.2 \\
& $m_{(\phi^{\pm})}=130$ GeV & 3.0 \\
\hline
$\phi^{\pm}\phi_{2}\rightarrow \tau E^{miss}_{T}$ & $m_{(\phi^{\pm})}=80$ GeV & 2.6 \\
\hline
$\phi^{\pm}\phi_{2}\rightarrow \mu E^{miss}_{T}$ & $m_{(\phi^{\pm})}=80$ GeV & 5.0 \\
\hline
\end{tabular}
\end{center}
\caption{Signal significance in different final states for the 7
TeV run, provided that 30 $fb^{-1}$ data is collected at this
energy before any switch to higher machine
energies.\label{7TeVsig}}
\end{table}
However with 30 $fb^{-1}$, as seen from Table \ref{7TeVsig},
 searching for the $\phi^{+}\phi^{-}$ process, a 5 $\sigma$ discovery for the $\mu\mu E^{miss}_{T}$ final state for $\phi^{\pm}$
 masses below $\sim 100$ GeV will be possible. A discovery for $\mu E^{miss}_{T}$ final state of $\phi^{\pm}\phi_{2}+\phi^{\pm}\delta_{2}$ is also possible for $m_{(\phi^{\pm})}=80$ GeV. The needed integrated luminosity for this conclusion is 30 $fb^{-1}$.

\section{Conclusions\label{con}}
In this paper, the possibility of discovering the SLIM model
proposed in \cite{yasaman} has been discussed. The focus has been
on search for a signal at the  LHC  in two different categories
based on the $\phi^{+}\phi^{-}$ and
$\phi^{\pm}\phi_{2}+\phi^{\pm}\delta_2$ production processes. In
the first category, two final states of $\tau~\mu$ and $\mu~\mu$
plus missing transverse energy have been analyzed in detail. The
main background processes have been identified and a cut based
search has been performed for the two final states separately. The
search focused on 30$fb^{-1}$ integrated luminosity at LHC. It has
been shown that the $\tau~\mu$ and $\mu~\mu$ final states would
have a signal significance of 2.8$\sigma$ and 9.2$\sigma$
respectively for $m_{(\phi^{\pm})}=80$ GeV. The $ee$ final state
has been estimated to have a significance of 4.1$\sigma$ based on
a simple rescaling of the $\mu\mu$ final state results. The above
search allows to have a signal significance exceeding 5$\sigma$ in
the $\mu\mu$ final state for the $\phi^{\pm}$ masses below 130 GeV
at 30$fb^{-1}$. In the second category, two final states of $\tau
E^{miss}_{T}$ and
 $\mu E^{miss}_{T}$ were analyzed in detail. Results indicate that a
 significance of 4.6$\sigma$ and 9.8$\sigma$ can be obtained for the
 above final states respectively. The $e E^{miss}_{T}$ final state is
 expected to have a significance of 6.5$\sigma$. A simple rescaling of
 the cross sections to their values at 7 TeV run of LHC shows that a
 discovery chance exists for $\mu\mu E^{miss}_{T}$ final state of
 $\phi^{+}\phi^{-}$ and also for $\mu E^{miss}_{T}$ final state of
 $\phi^{+}\phi_{2}$ even at this center of mass energy, provided that
 data corresponding to 30 $fb^{-1}$ integrated luminosity is collected
 before switching to higher machine energies. While the signal is
 observable at LHC,  because of the uncertainties in the background
 estimation, measuring the new couplings of the model is more challenging.
To make such a measurement a possibility, either these
uncertainties have to be reduced or a future collider with a
higher center of mass
 energy has to be employed.

{ Apart from the signals that we have discussed in the paper
($\phi^+ \phi^-$, $\phi^\pm \phi_2$ and $\phi^\pm \delta_2$), the
model can have other observable effects such as the
$\delta_2\phi_2$ production or invisible Higgs decay $H\to
\delta_1\delta_1$ \cite{yasaman}. Studying the prospect of
detecting all these modes is beyond the scope of the present paper
and will be presented elsewhere.}

\appendix \section{} In this appendix, we derive constraints on the
coupling $g_{i \alpha}$ from the neutrino mass matrix for the most
economic case with only two right-handed neutrinos. As discussed
in the text, in this case neutrino mass scheme is hierarchical.
\begin{itemize}
\item
\textit{Normal Hierarchical scheme:}
$$(m_\nu)_{\alpha \beta}=U_{PMNS}\cdot {\rm Diag}[0,\sqrt{\Delta
m_{sun}^2},\sqrt{\Delta m_{atm}^2}e^{i\xi}] \cdot U_{PMNS}^T$$
where $U_{PMNS}$ is the neutrino mixing matrix whose elements are
known up to the subleading $\theta_{13}$ effects.
However, we do not know the value of the Majorana phase $\xi$.

Using Eq.~(\ref{massMatrix}), we find that
\be \label{GiAlpha}g_{i\alpha}=\sum_j\frac{1}{A_i}( O^T\cdot {\rm Diag}[(\Delta
m_{sun}^2)^{1/4}, e^{i\xi/2}(\Delta
m_{atm}^2)^{1/4}])_{ij}(U_{PMNS})_{ \alpha j+1}\ ,\ee where $O$ is
an arbitrary orthogonal matrix: $O^T\cdot O=1$ so \be \label{O}
O=\left[\begin{matrix} \cos \theta & \sin \theta \cr -\sin \theta
& \cos \theta \end{matrix} \right].\ee In this case, we do not
have any prediction for the coupling ratios independent of the
arbitrary $\theta$.
\item
\textit{Inverted Hierarchical scheme:}
$$(m_\nu)_{\alpha \beta}=U_{PMNS}\cdot {\rm Diag}[\sqrt{\Delta m_{atm}^2},
\sqrt{\Delta m_{atm}^2+\Delta m_{sun}^2}e^{i\xi},0] \cdot
U_{PMNS}^T\ .$$

From Eq.~(\ref{massMatrix}), we therefore find
$$g_{i\alpha}=\sum_j\frac{1}{A_i}( O^T\cdot {\rm Diag}[(\Delta
m_{atm}^2)^{1/4}, e^{i\xi/2}(\Delta m_{atm}^2+ \Delta
m_{sun}^2)^{1/4}])_{ij}(U_{PMNS})_{ \alpha j}\ ,$$ where again $O$
is an arbitrary orthogonal matrix as in Eq. \ref{O}. It is
noteworthy that regardless of the values $\theta$ and $\xi$, \be
\frac{|g_{1 \tau}|^2}{|g_{1\mu}|^2}\simeq\frac{|g_{2
\tau}|^2}{|g_{2\mu}|^2}\simeq 1 +O(\theta_{13},
\theta_{23}-\pi/4).\ee However the values of ${|g_{1
e}|^2}/{|g_{1\tau}|^2}$ or ${|g_{2 e}|^2}/{|g_{2\tau}|^2}$ depend
on the unknown $\theta$ and $\xi$.

\end{itemize}
\


\end{document}